\documentclass[11pt]{article}
\usepackage{mathrsfs}
\usepackage{amsfonts}
\usepackage{amsmath}
\usepackage{amssymb,amsthm}
\usepackage{hyperref,color}
\usepackage{exscale,relsize}

\numberwithin{equation}{section}
\allowdisplaybreaks[4]

\author{ Yu Hou   \and  Engui Fan\footnote{Corresponding
author and  e-mail address:
      faneg@fudan.edu.cn}}
\date{   \small{ School of Mathematical Sciences, Shanghai Center for Mathematical Sciences \\
 and Key Laboratory of Mathematics for Nonlinear Science, \\ Fudan
University, Shanghai 200433, P.R. China}}
\title{\bf \Large{Algebro-geometric solutions for the two-component Hunter-Saxton hierarchy} }
\begin{document}
\maketitle

\begin{abstract}
This paper is dedicated to provide theta function representations of
algebro-geometric solutions and related crucial quantities for the
two-component Hunter-Saxton (HS2) hierarchy through studying an algebro-geometric
initial value problem. Our main tools include the polynomial
recursive formalism, the hyperelliptic
curve with finite number of genus, the Baker-Akhiezer functions, the
meromorphic function, the Dubrovin-type equations for auxiliary
divisors, and the associated trace formulas. With the help of these
tools, the explicit representations of the algebro-geometric solutions  are
obtained for the entire HS2 hierarchy.
\end{abstract}

\section{Introduction}

In this paper, we consider the following integrable two-component Hunter-Saxton (HS2)
system:
 \begin{equation}\label{1.1}
  \left\{
    \begin{array}{ll}
      m_t+2u_xm+um_x+\sigma\rho \rho_x=0,  \\
      \rho_t+(u\rho)_x=0,
    \end{array}
  \right.
 \end{equation}
where $m=-u_{xx}$, $\sigma=\pm 1$,
which was recently introduced by Constantin and Ivanov in \cite{Constantin01}.
The variable $u(x, t)$ can be interpreted as
the horizontal fluid velocity and the variable $\rho (x, t)$ describes
the horizontal deviation of the surface from equilibrium,
all measured in dimensionless units \cite{Constantin01}.

The HS2 system arises in the short-wave (or high-frequency) limits,
obtained via the space-time scaling
$   (x,t)\mapsto (\varepsilon x, \varepsilon t)   $
and letting $\varepsilon$ tend to zero in the resulting equation,
of the two-component integrable Camassa-Holm  system \cite{Constantin01, Escher, Olver, Wu}.
 This system, reading as
(\ref{1.1}) with $m$ replaced by $(1-\partial_{xx}^2)u$,
was derived from the Green-Naghdi equations, which are approximations to the governing
equations for water waves. It has recently been the object of intensive study,
see, e.g., \cite{chen, Constantin01, Gui, Holm001, zhangp}.
The  HS2 system (\ref{1.1}) is integrable, it has Lax pair  \cite{Constantin01} and
a bi-Hamiltonian structure \cite{Constantin01, Olver}.
It is also a particular
case of the Gurevich-Zybin system describing the dynamics in a model of nondissipative
dark matter \cite{chuanx2, Pavlov, Wunsch}.
The mathematical properties of Eq.(\ref{1.1}) have been studied further in many works,
see, e.g., \cite{Constantin01, chuanx1, chuanx2, liu1, song, Wu, Wunsch}.

For $\rho=0$, the system (\ref{1.1}) becomes the Hunter-Saxton (HS) equation \cite{Hunter},
which models the propagation of weakly nonlinear orientation waves in a massive
nematic liquid crystal director field.
Here, $u(x,t)$ stands for the director field of a nematic liquid crystal,
$x$ is a
space variable in a reference frame moving with the linearized wave velocity,
and $t$ is a slow time variable.
The field of unit vectors $( \cos {u(x,t)}, \sin {u(x,t)})$ describes
the orientation of the molecules \cite{chuanx1, Hunter, Wunsch}.

The HS equation also describes the high-frequency limit \cite{Dai1, Hunter1} of the Camassa-Holm (CH) equation
--a model equation for shallow
water waves \cite{Camassa01, Constantin02, Johnson}
and a re-expression of the geodesic flow on the diffeomorphism group of
the circle \cite{Constantin08}  with a bi-Hamiltonian structure \cite{Fokas} which is completely integrable \cite{C1}.
The HS equation is also a completely integrable system with a bi-Hamiltonian structure, and hence, it possesses
a Lax pair, an infinite family of commuting Hamiltonian flows, as well as an associated sequence of
conservation laws \cite{BSS, Hunter, Hunter1, Olver, Reyes}. It also describes
the geodesic flow on the homogeneous space related to the Virasoro group \cite{Khesin, Lenells1, Lenells2}.
Recently, algebro-geometric solutions for the HS hierarchy was investigated in \cite{18a}.
Moreover, peakon solutions, global weak solutions and the Cauchy problem of the system
(\ref{1.1}) were discussed in \cite{Constantin01, chuanx1, liu1, Wunsch}.
However, within the knowledge of the authors, the
algebro-geometric solutions of the entire HS2 hierarchy are not studied yet.

 The principal subject of this paper concerns algebro-geometric quasi-periodic solutions
 of the whole HS2 hierarchy, of which (\ref{1.1}) ($\sigma=1$) is just the first of infinitely many members.
 Algebro-geometric solution, as an important feature of integrable system,
 is a kind of explicit solution closely related to the inverse spectral
 theory \cite{4, 7}, \cite{9}-\cite{11}. In a degenerated case of the
 algebro-geometric solution, the multi-soliton solution and periodic solution
 in elliptic function type may be obtained \cite{7, 33, 8}.
 A systematic approach, proposed by Gesztesy and Holden to construct
 algebro-geometric solutions for integrable equations,   has been extended to
 the whole (1+1) dimensional integrable hierarchy, such as the AKNS hierarchy,
 the CH hierarchy, etc. \cite{13}-\cite{12}.
 Recently, we investigated algebro-geometric solutions for the
 modified CH hierarchy and the Degasperis-Procesi hierarchy \cite{18, 17}.

 The outline of the present paper is as follows.

 In section 2, based
 on the polynomial recursion formalism, we derive the HS2 hierarchy,
 associated with the $2 \times 2$ spectral problem. A hyperelliptic curve
 $\mathcal{K}_{n}$ of arithmetic genus $n$ is introduced with
 the help of the characteristic polynomial of Lax matrix $V_n$ for the
 stationary HS2 hierarchy.

 In Section 3,  we decompose the stationary
 HS2 equations into a system of Dubrovin-type
 equations. Moreover, we obtain the stationary trace formulas
 for the HS2 hierarchy.

 In Section 4, we present the first set of our results, the explicit
 theta function representations of the potentials $u, \rho$
 for the entire stationary HS2 hierarchy. Furthermore, we study the
 initial value problem on an algebro-geometric curve for the stationary
 HS2 hierarchy.

 In Sections 5 and 6, we extend the analyses of Sections 3 and 4, respectively,
 to the time-dependent case. Each equation in the HS2
 hierarchy is permitted to evolve in terms of an independent time
 parameter $t_r$. As  initial data, we use a stationary solution of
 the $n$th equation and then construct a time-dependent solution of
 the $r$th equation of the HS2 hierarchy.
 The Baker-Akhiezer function,  the analogs
 of the Dubrovin-type equations, the trace formulas, and the theta function
 representations in Section 4 are all extended to the time-dependent
 case.

 Finally, we remark that although our focus in this paper is on Eq.(\ref{1.1}) with $\sigma=1$,
 all of the arguments presented here can be adapted, without obvious modifications,
 to study the corresponding equation with $\sigma=-1$.

\section{The HS2 hierarchy}
 In this section, we provide the construction of  HS2 hierarchy and derive the
 corresponding
 sequence of zero-curvature pairs using a polynomial recursion formalism.
 Moreover, we introduce the underlying hyperelliptic curve in connection with
 the stationary HS2 hierarchy.

 Throughout this section, we make the following hypothesis.

\newtheorem{hyp1}{Hypothesis}[section]
 \begin{hyp1}
      In the stationary case, we assume that
  \begin{equation}\label{2.1}
    \begin{split}
    & u, \rho \in C^\infty(\mathbb{R}),
   ~\partial_x^k u, \partial_x^k \rho  \in L^\infty (\mathbb{R}), ~ k\in \mathbb{N}_0.
    \end{split}
  \end{equation}
    In the time-dependent case, we suppose
  \begin{equation}
    \begin{split}\label{2.2}
    & u(\cdot,t), \rho(\cdot,t) \in C^\infty (\mathbb{R}),\ \
     \partial_x^ku(\cdot,t), \partial_x^k \rho(\cdot,t) \in L^\infty
    (\mathbb{R}),~~ k\in \mathbb{N}_0,~ t\in \mathbb{R},\\
    & u(x,\cdot), u_{xx}(x,\cdot), \rho(x,\cdot), \rho_x(x, \cdot) \in C^1(\mathbb{R}),
       \quad x\in \mathbb{R}.\\
  \end{split}
  \end{equation}
\end{hyp1}

We first introduce the basic polynomial recursion formalism. Define
 $\{f_{ l}\}_{l\in\mathbb{N}_{0}}$,
 $\{g_{ l}\}_{l\in\mathbb{N}_{0}}$, and
 $\{h_{ l}\}_{l\in\mathbb{N}_{0}}$
recursively by
 \begin{equation}\label{2.3}
      \begin{split}
        & f_0=\frac{1}{2}, \\
       & f_{l,x}=-2\mathcal{G}\left(2\rho^2f_{l-2,x}+2u_{xx}f_{l-1,x}+2\rho\rho_xf_{l-2}+u_{xxx}f_{l-1}\right),
        \quad l \in \mathbb{N},\\
      & g_l=\frac{1}{2}f_{l+1,x},\quad l \in \mathbb{N}_0,\\
      & h_l=-g_{l+1,x}-\rho^2f_{l}-u_{xx}f_{l+1}, \quad l \in \mathbb{N}_0,\\
      \end{split}
 \end{equation}
where $\mathcal{G}$ is given by
    \begin{equation}\label{2.4}
     \begin{split}
     & \mathcal{G}: L^\infty(\mathbb{R}) \rightarrow
        L^\infty(\mathbb{R}),\\
     & (\mathcal{G}v)(x)=\int_{-\infty}^x \int_{-\infty}^{x_1}
     v(y) ~dy dx_1 , \quad x\in\mathbb{R},~
        v\in L^\infty(\mathbb{R}).
      \end{split}
    \end{equation}
One observes that $\mathcal{G}$ is the resolvent of the
one-dimensional Laplacian operator, that is,
    \begin{equation}\label{2.5}
        \mathcal{G}=\Big( \frac{d^2}{dx^2} \Big)^{-1}.
    \end{equation}
Explicitly, one computes
   \begin{equation}\label{2.6}
     \begin{split}
      & f_0=\frac{1}{2},\\
      & f_1=-u+c_1,\\
      & f_2=\mathcal{G}(u_x^2-\rho^2+2uu_{xx})+2c_1(-u) +c_2,\\
      & g_0=-\frac{1}{2}u_x,\\
      & g_1=\frac{1}{2} \mathcal{G}(4u_xu_{xx}+2uu_{xxx}-2\rho\rho_{x})+2c_1(-\frac{1}{2}u_x),\\
      & h_0=-\frac{1}{2}f_{2,xx}-\rho^2f_0-u_{xx}f_1, ~\mathrm{etc}.,
     \end{split}
   \end{equation}
where $\{c_l\}_{l\in\mathbb{N}}\subset\mathbb{C}$ are integration
constants.

Next, it is convenient to introduce the corresponding homogeneous coefficients
$\hat{f}_{l}, \hat{g}_{l},$ and  $\hat{h}_{l},$ defined by the vanishing of the
integration constants $c_k,k=1,\ldots,l,$
 \begin{equation}\label{2.8}
       \begin{split}
        & \hat{f}_0=f_0=\frac{1}{2}, \quad \hat{f}_{l}=f_{l}|_{c_k=0,~k=1,\ldots,l},
         \quad l \in \mathbb{N},
        \quad \\
        & \hat{g}_0=g_0=-\frac{1}{2}u_x, \quad
        \hat{g}_{l}=g_{l}|_{c_k=0,~k=1,\ldots,l},
        \quad  l \in \mathbb{N},\\
        &   \hat{h}_0=h_0, \quad
           \quad \hat{h}_{l}=h_{l}|_{c_k=0,~k=1,\ldots,l},
           \quad l \in \mathbb{N}.
       \end{split}
    \end{equation}
Hence,
    \begin{equation}\label{2.9}
      f_{l}=\sum_{k=0}^{l}2c_{l-k}\hat{f}_{k}, \quad
      g_{l}=\sum_{k=0}^{l}2c_{l-k}\hat{g}_{k}, \quad
      h_{l}=\sum_{k=0}^{l}2c_{l-k}\hat{h}_{k}, \quad
       l\in \mathbb{N}_0,
    \end{equation}
defining
    \begin{equation}\label{2.10}
        c_0=\frac{1}{2}.
    \end{equation}
Now, given Hypothesis 2.1,
one introduces the following $2\times 2$ matrix $U$ by
     \begin{equation}\label{2.11}
       \psi_x=U(z,x)\psi=
       \left(
         \begin{array}{cc}
           0 & 1 \\
           -z^{-2}\rho^2-z^{-1}u_{xx} & 0 \\
         \end{array}
       \right)
       \psi,
     \end{equation}
and for each $n\in\mathbb{N}_0,$ the following $2\times 2$
matrix $V_n$ by
\begin{equation} \label{2.12}
       \psi_{t_n}=V_n(z)\psi,
    \end{equation}
  with
    \begin{equation}\label{2.13}
      V_n(z)=
      \left(
        \begin{array}{cc}
          -G_n(z) & F_{n}(z) \\
          z^{-2}H_n(z) & G_n(z) \\
        \end{array}
      \right),
    \quad z \in \mathbb{C} \setminus \{0\},
    \quad n\in\mathbb{N}_0,
    \end{equation}
assuming $F_{n}$, $G_n$, and $H_n$ to be polynomials\footnote{$F_{n}, G_n, H_n$ are
polynomials of degree $n+1,n,n+1$, respectively.}
with respect to $z$ and $C^{\infty}$ in $x$. The compatibility
condition of linear system (\ref{2.11}) and (\ref{2.12}) yields the
stationary zero-curvature equation
\begin{equation}\label{2.14}
   -V_{n,x}+[U,V_n]=0,
\end{equation}
which is equivalent to
   \begin{align}
      & F_{n,x}= 2G_n, \label{2.15a}\\
      & H_{n,x}= 2(\rho^2+zu_{xx})G_n, \label{2.15b}\\
      & z^2G_{n,x}=-H_n- (\rho^2+zu_{xx})F_{n}.\label{2.15c}
    \end{align}
From (\ref{2.15a})-(\ref{2.15c}), one infers that
   \begin{equation}\label{2.18}
     \frac{d}{dx} \mathrm{det} (V_n(z,x))=
     -\frac{1}{z^2} \frac{d}{dx} \Big(
     z^2G_n(z,x)^2+F_{n}(z,x)H_n(z,x)
     \Big)=0,
   \end{equation}
and hence
   \begin{equation}\label{2.19}
    z^2G_n(z,x)^2+F_{n}(z,x)H_n(z,x)=R_{2n+2}(z),
   \end{equation}
where the polynomial $R_{2n+2}$ of degree $2n+2$ is $x$-independent.
In another way, one can write
$R_{2n+2}$ as
\begin{equation}\label{2.20}
   R_{2n+2}(z)=\left(\frac{1}{4}u_x^2+\frac{1}{2}h_0\right)\prod_{m=0}^{2n+1}(z-E_m),\quad
   \{E_m\}_{m=0,\ldots,2n+1}\in\mathbb{C}.
   \end{equation}
Here, we emphasize that the coefficient $(\frac{1}{4}u_x^2+\frac{1}{2}h_0)$ is a
constant. In fact, equation (\ref{2.18}) is equivalent to
    \begin{equation}\label{2.21}
     2z^2G_nG_{n,x}+F_{n}H_{n,x}+H_nF_{n,x}=0.
    \end{equation}
Then comparing the coefficient of powers $z^{2n+2}$ yields
   \begin{equation}\label{2.22}
    2g_0g_{0,x}+f_0h_{0,x}+h_0f_{0,x}=0,
   \end{equation}
which indicates
    \begin{equation}\label{2.23}
        \frac{1}{2}u_xu_{xx}+\frac{1}{2}h_{0,x}=0.
    \end{equation}
Therefore,
    \begin{equation}\label{2.24}
        \frac{1}{4}u_x^2+\frac{1}{2}h_0=\partial^{-1} \left(\frac{1}{2}u_xu_{xx}+\frac{1}{2}h_{0,x}\right)=
          \mathrm{constant}.
    \end{equation}
For simplicity, we denote it
     by $a^2$,
    $a\in\mathbb{C}.$
  Then, $R_{2n+2}(z)$ can be
rewritten as
    \begin{equation}\label{2.25}
   R_{2n+2}(z)=a^2 \prod_{m=0}^{2n+1}(z-E_m),\quad
   \{E_m\}_{m=0,\ldots,2n+1}\in\mathbb{C}.
   \end{equation}

Next, we compute
the characteristic polynomial $\mathrm{det}(yI-z V_n)$ of Lax matrix
$z V_n$,
    \begin{align}\label{2.26}
      \mathrm{det}(yI-z V_n)&=
        y^2-z^2G_n(z)^2-F_{n}(z)H_n(z)
         \nonumber \\
      &= y^2-R_{2n+2}(z)=0,
    \end{align}
and then introduce the (possibly singular) hyperelliptic curve
$\mathcal{K}_n$ of arithmetic genus $n$ defined by
\begin{equation}\label{2.27}
     \mathcal {K}_n:\mathcal {F}_n(z,y)=y^2-R_{2n+2}(z)=0.
   \end{equation}
 In the following, we will occasionally impose further constraints on
 the zeros $E_m$ of $R_{2n+2}$ introduced in (\ref{2.25}) and assume
 that
 \begin{equation}\label{2.26a}
  E_m\in\mathbb{C}, \quad E_{m}\neq E_{m^\prime},\quad
 \forall m\neq m^\prime,\quad  m,m^\prime=0,\ldots,2n+1.
 \end{equation}

The stationary zero-curvature equation
(\ref{2.14}) implies polynomial recursion relations (\ref{2.3}).
Introducing the following polynomials $F_{n}(z), G_n(z)$, and
$H_n(z)$ with respect to the spectral parameter $z$,
   \begin{align}
    & F_{n}(z)=\sum_{l=0}^{n+1} f_{l} z^{n+1-l},\label{2.28}
         \\
    &
       G_n(z)=\sum_{l=0}^{n} g_{l} z^{n-l}, \label{2.29}
         \\
    &
      H_n(z)=\sum_{l=0}^{n+1} h_{l} z^{n+1-l}. \label{2.30}
   \end{align}
Inserting (\ref{2.28})-(\ref{2.30}) into (\ref{2.15a})-(\ref{2.15c}) then yields
the recursion relations (\ref{2.3}) for $f_{l},$ $l=0,\ldots, n+1,$
and $g_{l}$, $l=0,\ldots, n.$  For fixed $n \in \mathbb{N}_0$,
we obtain the recursion relations for $h_{l}, $ $l=0, \ldots,
n-1$ in (\ref{2.3}) and
    \begin{equation}\label{2.31}
         h_{n}=-\rho^2f_{n}-u_{xx}f_{n+1}, \quad
           h_{n+1}=-\rho^2f_{n+1}.
    \end{equation}
Moreover, from (\ref{2.15b}), one infers that
   \begin{equation}\label{2.32}
     \begin{split}
    &-h_{n,x}+\rho^2f_{n,x}+u_{xx}f_{n+1,x}=0, \quad n\in \mathbb{N}_0,\\
    &   -h_{n+1,x}+\rho^2f_{n+1,x}=0, \quad n\in \mathbb{N}_0.\\
     \end{split}
   \end{equation}
Then using  (\ref{2.31}) and (\ref{2.32})
permits one to write the stationary  HS2 hierarchy as
 \begin{equation}\label{2.34}
        \textrm{s-HS2}_n(u,\rho)=
        \left(
          \begin{array}{c}
            2u_{xx}f_{n+1,x}+u_{xxx}f_{n+1}+2\rho\rho_xf_{n}+2\rho^2f_{n,x} \\
            -2\rho\rho_xf_{n+1}-2\rho^2f_{n+1,x} \\
          \end{array}
        \right)=0,
        \quad n\in \mathbb{N}_0.
      \end{equation}
We record the first equation explicitly,
    \begin{equation}\label{2.35}
      \begin{split}
       & \textrm{s-HS2}_0(u,\rho)=
        \left(
         \begin{array}{c}
           -2u_xu_{xx}-uu_{xxx}+\rho\rho_x+c_1u_{xxx} \\
           \rho_x u+\rho u_x-c_1\rho_x \\
         \end{array}
       \right)=0.
       \end{split}
    \end{equation}

By definition, the set of solutions of (\ref{2.34}) represents the
class of algebro-geometric HS2 solutions, with $n$ ranging in
$\mathbb{N}_0$ and $c_l$ in $\mathbb{C},~l\in\mathbb{N}$. We call
the stationary algebro-geometric HS2 solutions $u,$ $\rho$ as HS2 potentials
at times.

\newtheorem{rem2.2}[hyp1]{Remark}
  \begin{rem2.2}
    Here, we emphasize that if $u$, $\rho$ satisfy one of the stationary HS$2$ equations
    in
    $(\ref{2.34})$ for a particular value of $n$, then they satisfy infinitely many such equations
    of order higher than $n$ for certain choices of integration
    constants $c_l$. This is a common characteristic of the general
    integrable soliton equations such as the KdV, AKNS, and CH
    hierarchies \cite{15}.
  \end{rem2.2}

Next, we introduce the corresponding homogeneous polynomials
$\widehat{F}_{l}, \widehat{G}_{l}, \widehat{H}_{l}$ by
    \begin{eqnarray}
    &&
      \widehat{F}_{l}(z)=F_{l}(z)|_{c_k=0,~k=1,\dots,l}
      =\sum_{k=0}^{l} \hat{f}_{k} z^{l-k},
      ~~ l=0,\ldots,n+1, \label{2.36a1}\\
    &&
      \widehat{G}_{l}(z)=G_{l}(z)|_{c_k=0,~k=1,\dots,l}
      =\sum_{k=0}^l \hat{g}_{k} z^{l-k},
      ~~ l=0,\ldots,n,\\
    &&
      \widehat{H}_{l}(z)=H_{l}(z)|_{c_k=0,~k=1,\dots,l}
      =\sum_{k=0}^l \hat{h}_{k} z^{l-k},
      ~~ l=0,\ldots,n-1,\\
    &&
      \widehat{H}_{n}(z)= -\rho^2\hat{f}_{n}-u_{xx}\hat{f}_{n+1}+
      \sum_{k=0}^{n-1} \hat{h}_{k} z^{n-k}, \\
    &&
       \widehat{H}_{n+1}(z)= -\rho^2\hat{f}_{n+1}+ z\widehat{H}_{n}(z).\label{2.36a2}
          \end{eqnarray}

In accordance with our notation introduced in (\ref{2.8}) and (\ref{2.36a1})-(\ref{2.36a2}),
the corresponding homogeneous stationary HS2
 equations are then defined by
      \begin{equation}\label{2.40}
        \textrm{s-}\widehat{\mathrm{HS2}}_n(u,\rho)=
        \textrm{s-HS2}_n(u,\rho)|_{c_l=0,~l=1,\dots,n}=0,
        \quad n\in\mathbb{N}_0.
      \end{equation}

At the end of this section, we turn to the time-dependent HS2 hierarchy. In this case, $u$, $\rho$
are considered as  functions of both space and time. We introduce
a deformation parameter $t_n \in \mathbb{R}$ in $u$ and $\rho$, replacing
$u(x),\rho(x)$ by $u(x,t_n), \rho(x,t_n)$, for each equation in the hierarchy. In
addition, the definitions (\ref{2.11}), (\ref{2.13}),
and (\ref{2.28})-(\ref{2.30}) of $U,$ $V_n$ and $F_{n}, G_n$, and $H_n,$ respectively,
still apply. The corresponding
zero-curvature equation reads
 \begin{equation}\label{2.41}
   U_{t_n}-V_{n,x}+[U,V_n]=0, \quad n\in \mathbb{N}_0,
   \end{equation}
which results in the following set of equations
    \begin{eqnarray}
      && F_{n,x}=2G_n, \label{2.42c1} \\
      && z^2G_{n,x}=-H_n -(\rho^2+zu_{xx})F_{n}, \label{2.42c2}\\
      && -2\rho\rho_{t_n}-zu_{xxt_n}-H_{n,x}+
      2(\rho^2+zu_{xx})G_n=0.  \label{2.42c3}
    \end{eqnarray}
For fixed $n \in \mathbb{N}_0$, inserting the polynomial
expressions for $F_{n}$, $G_n$, and $H_n$ into (\ref{2.42c1})-(\ref{2.42c3}),
respectively, first yields recursion relations (\ref{2.3}) for
$f_{l}|_{l=0,\ldots,n+1}$, $g_{l}|_{l=0,\ldots,n}$,
$h_{l}|_{l=0,\ldots,n-1}$ and
      \begin{equation}\label{2.45}
        h_{n}=-\rho^2f_{n}-u_{xx}f_{n+1}, \quad
        h_{n+1}=-\rho^2f_{n+1}.
      \end{equation}
Moreover, using (\ref{2.42c3}), one finds
     \begin{equation}\label{2.46}
       \begin{split}
       & -u_{xxt_n}-h_{n,x}+\rho^2f_{n,x}+u_{xx}f_{n+1,x}=0,
        \quad n\in \mathbb{N}_0, \\
       & 2\rho\rho_{t_n}+h_{n+1,x}-\rho^2f_{n+1,x}=0, \quad n\in \mathbb{N}_0.
       \end{split}
     \end{equation}
Hence, using (\ref{2.45}) and (\ref{2.46})
permits one to write the time-dependent  HS2 hierarchy as
\begin{equation}\label{2.48}
        \textrm{HS2}_n(u,\rho)=
        \left(
          \begin{array}{c}
            -u_{xxt_n} +2u_{xx}f_{n+1,x}+u_{xxx}f_{n+1}
            +2\rho\rho_xf_{n}+2\rho^2f_{n,x} \\
            2\rho\rho_{t_n}-2\rho\rho_xf_{n+1}-2\rho^2f_{n+1,x} \\
          \end{array}
        \right)=0, \quad n\in \mathbb{N}_0.
      \end{equation}
For convenience, we record the first equation in this hierarchy explicitly,
     \begin{equation}\label{2.49}
      \begin{split}
       &
       \mathrm{HS2}_0(u,\rho)=
       \left(
         \begin{array}{c}
           -u_{xxt_0}-2u_xu_{xx}-uu_{xxx}+\rho\rho_x+c_1u_{xxx} \\
           \rho_{t_0}+\rho_x u +\rho u_x-c_1\rho_x \\
         \end{array}
       \right)=0.
       \end{split}
     \end{equation}
The first equation $\mathrm{HS2}_0(u,\rho)=0$ (with $c_1=0$) in the
hierarchy represents the  HS2 system as discussed in section
1. Similarly, one can introduce the corresponding homogeneous HS2
hierarchy by
    \begin{equation}\label{2.50}
        \widehat{\mathrm{HS2}}_n(u,\rho)=\mathrm{HS2}_n(u,\rho)|_{c_l=0,~l=1,\ldots,n}=0,
        \quad n\in\mathbb{N}_0.
    \end{equation}

In fact, since the Lenard recursion formalism is almost universally
adopted in the contemporary literature, we thought it might be
worthwhile to use the Gesztesy's method, the polynomial recursion formalism,
to construct the HS2 hierarchy.

\section{The stationary HS2 formalism}
 This section is devoted to a detailed study of the stationary HS2 hierarchy.
 We first
 define a fundamental meromorphic function $\phi(P,x)$ on the hyperelliptic
 curve $\mathcal{K}_n$, using the polynomial recursion formalism described in section 2,
 and then study the properties of the
 Baker-Akhiezer function $\psi(P,x,x_0)$, Dubrovin-type equations,
 and trace formulas.

For major parts of this section, we assume (\ref{2.1}), (\ref{2.3}),
(\ref{2.6}), (\ref{2.11})-(\ref{2.15c}), (\ref{2.27})-(\ref{2.30}),
and (\ref{2.34}), keeping $n\in\mathbb{N}_0$
fixed.

Recall the hyperelliptic curve $\mathcal{K}_n$
     \begin{equation}\label{3.1}
       \begin{split}
        & \mathcal{K}_n:  \mathcal{F}_n(z,y)=y^2-R_{2n+2}(z)=0, \\
        & R_{2n+2}(z)=a^2 \prod_{m=0}^{2n+1} (z-E_m), \quad
      \{E_m\}_{m=0,\ldots,2n+1} \in \mathbb{C},
      \end{split}
     \end{equation}
which is compactified by joining two points at infinity
$P_{\infty_\pm}$, with $P_{\infty_+} \neq P_{\infty_-}$. But for
notational simplicity, the compactification is also denoted by
$\mathcal{K}_n$. Hence, $\mathcal{K}_n$ becomes a two-sheeted Riemann surface of
arithmetic genus $n$. Points $P$ on
      $\mathcal{K}_{n}\backslash\{P_{\infty\pm}\}$
are denoted by $P=(z,y(P))$, where $y(\cdot)$ is the
meromorphic function on $\mathcal{K}_{n}$ satisfying
       $\mathcal{F}_n(z,y(P))=0.$

The complex structure on $\mathcal{K}_{n}$ is defined in the usual
way by introducing local coordinates
$$\zeta_{Q_0}:P\rightarrow(z-z_0)$$
near points $Q_0=(z_0,y(Q_0))\in \mathcal{K}_{n},$
which are neither branch nor singular points of
$\mathcal{K}_{n}$; near the branch and singular points $Q_1=(z_1,y(Q_1)) \in \mathcal{K}_{n}$, the local coordinates are
   $$\zeta_{Q_1}:P \rightarrow (z-z_1)^{1/2};$$
near the points $P_{\infty_\pm} \in \mathcal{K}_{n}$, the local
coordinates are
   $$\zeta_{P_{\infty_\pm}}:P \rightarrow z^{-1}.$$

The holomorphic map
   $\ast,$ changing sheets, is defined by
       \begin{eqnarray}\label{3.2}
       && \ast: \begin{cases}
                        \mathcal{K}_{n}\rightarrow\mathcal{K}_{n},
                       \\
                       P=(z,y_j(z))\rightarrow
                       P^\ast=(z,y_{j+1(\mathrm{mod}~
                       2)}(z)), \quad j=0,1,
                      \end {cases}
                     \nonumber \\
      && P^{\ast \ast}:=(P^\ast)^\ast, \quad \mathrm{etc}.,
       \end{eqnarray}
where $y_j(z),\, j=0,1$ denote the two branches of $y(P)$ satisfying
$\mathcal{F}_{n}(z,y)=0$,  namely,
        \begin{equation}\label{3.17}
          (y-y_0(z))(y-y_1(z))
          =y^2-R_{2n+2}(z)=0.
        \end{equation}
 Taking into account (\ref{3.17}), one easily finds
        \begin{equation}\label{3.18}
           \begin{split}
             & y_0+y_1=0,\\
             & y_0y_1=-R_{2n+2}(z),\\
             & y_0^2+y_1^2=2R_{2n+2}(z).\\
           \end{split}
        \end{equation}\\
Moreover, positive divisors on
$\mathcal{K}_{n}$ of degree $n$ are denoted by
        \begin{equation}\label{3.3}
          \mathcal{D}_{P_1,\ldots,P_{n}}:
             \begin{cases}
              \mathcal{K}_{n}\rightarrow \mathbb{N}_0,\\
              P\rightarrow \mathcal{D}_{P_1,\ldots,P_{n}}=
                \begin{cases}
                  \textrm{ $k$ \quad if $P$ occurs $k$
                      times in $\{P_1,\ldots,P_{n}\},$}\\
                   \textrm{ $0$ \quad if $P \notin
                     $$ \{P_1,\ldots,P_{n}\}.$}
                \end{cases}
             \end{cases}
        \end{equation}

Next, we define the stationary Baker-Akhiezer function
$\psi(P,x,x_0)$ on $\mathcal{K}_{n}\setminus \{P_{\infty_+},
P_{\infty_-},P_0=(0, i \rho f_{n+1})\}$ by
       \begin{equation}\label{3.4}
         \begin{split}
          & \psi(P,x,x_0)=\left(
                            \begin{array}{c}
                              \psi_1(P,x,x_0) \\
                              \psi_2(P,x,x_0) \\
                            \end{array}
                          \right), \\
          & \psi_x(P,x,x_0)=U(u(x),\rho(x),z(P))\psi(P,x,x_0),\\
           & zV_n(u(x),\rho(x),z(P))\psi(P,x,x_0)=y(P)\psi(P,x,x_0),\\
          & \psi_1(P,x_0,x_0)=1; \\
           &   P=(z,y)\in \mathcal{K}_{n}
           \setminus \{P_{\infty_+},P_{\infty_-},P_0=(0,i \rho f_{n+1})\},~(x,x_0)\in \mathbb{R}^2.
         \end{split}
       \end{equation}
Closely related to $\psi(P,x,x_0)$ is the following meromorphic
function $\phi(P,x)$ on $\mathcal{K}_{n}$ defined by
       \begin{equation}\label{3.5}
         \phi(P,x)= z
         \frac{ \psi_{1,x}(P,x,x_0)}{\psi_1(P,x,x_0)},
         \quad P\in \mathcal{K}_{n},~ x\in \mathbb{R}
       \end{equation}
such that
    \begin{equation}\label{3.6}
    \psi_1(P,x,x_0)=\mathrm{exp}\left(z^{-1}\int_{x_0}^x
         \phi(P,x^\prime)~ dx^\prime
         \right),
         \quad P\in \mathcal{K}_{n}\setminus \{P_{\infty_+}, P_{\infty_-},
         P_0\}.
    \end{equation}
Then, based on (\ref{3.4}) and (\ref{3.5}), a direct calculation
shows that
    \begin{align}\label{3.7}
        \phi(P,x)&=\frac{y+zG_n(z,x)}{F_{n}(z,x)}
           \nonumber \\
        &=
         \frac{ H_n(z,x)}{y-zG_n(z,x)},
    \end{align}
and
      \begin{equation}\label{3.8}
        \psi_2(P,x,x_0)= \psi_1(P,x,x_0)\phi(P,x)/z.
      \end{equation}

In the following, the roots of polynomials $F_{n}$ and $H_n$ will
play a special role, and hence, we introduce on $\mathbb{C}\times\mathbb{R}$
 \begin{equation}\label{3.9}
         F_{n}(z,x)=\frac{1}{2}\prod_{j=0}^{n}(z-\mu_j(x)),
            \quad
            H_n(z,x)=h_0\prod_{l=0}^{n}(z-\nu_l(x)).
         \end{equation}
Moreover, we introduce
      \begin{equation}\label{3.10}
        \hat{\mu}_j(x)
            =(\mu_j(x),-\mu_j(x)G_n(\mu_j(x),x))
            \in \mathcal{K}_{n}, ~
            j=0,\ldots,n,~x\in\mathbb{R},
      \end{equation}
and
       \begin{equation}\label{3.11}
        \hat{\nu}_l(x)
            =(\nu_l(x),\nu_l(x)G_n(\nu_l(x),x))
            \in \mathcal{K}_{n}, ~
            l=0,\ldots,n,~x\in\mathbb{R}.
      \end{equation}
Due to assumption (\ref{2.1}), $u$ and $\rho$ are smooth and bounded, and hence,
$F_{n}(z,x)$ and $H_n(z,x)$ share the same property. Thus, one
concludes
    \begin{equation}\label{3.12}
        \mu_j,\nu_l \in C(\mathbb{R}), \quad
        j,l=0,\dots,n,
    \end{equation}
taking multiplicities (and appropriate reordering)
of the zeros of $F_{n}$ and $H_n$ into account.
From (\ref{3.7}),
the divisor $(\phi(P,x))$ of $\phi(P,x)$ is
given by
         \begin{equation}\label{3.14}
           (\phi(P,x))=\mathcal{D}_{\hat{\nu}_0(x)  \underline{\hat{\nu}}(x)}(P)
           -\mathcal{D}_{\hat{\mu}_0(x) \underline{\hat{\mu}}(x)}(P).
         \end{equation}
Here, we abbreviated
     \begin{equation}\label{3.15}
     \underline{\hat{\mu}}=\{\hat{\mu}_1,\ldots,\hat{\mu}_n\},
      \quad
      \underline{\hat{\nu}}=\{\hat{\nu}_1,\ldots,\hat{\nu}_{n}\}
      \in \mathrm{Sym}^n(\mathcal{K}_n).
     \end{equation}

Further properties of $\phi(P,x)$ are summarized as follows.
\newtheorem{lem3.1}{Lemma}[section]
 \begin{lem3.1}\label{lemma3.1}
    Suppose $(\ref{2.1})$, assume the $n$th stationary $HS2$ equation $(\ref{2.34})$ holds, and
    let
    $P=(z,y)\in \mathcal{K}_{n}\setminus \{P_{\infty_+}, P_{\infty_-},P_0\},$
    $(x,x_0) \in \mathbb{R}^2$. Then $\phi$ satisfies the Riccati-type equation
      \begin{equation}\label{3.19}
       \phi_x(P)+z^{-1}\phi(P)^2=-z^{-1}\rho^2-u_{xx},
      \end{equation}
      as well as
       \begin{align}
        & \phi(P)\phi(P^\ast)=-\frac{H_n(z)}{F_{n}(z)},\label{3.20}
          \\
        &
         \phi(P)+\phi(P^\ast)=\frac{2zG_n(z)}{F_{n}(z)}, \label{3.21}
          \\
        &
        \phi(P)-\phi(P^\ast)=\frac{2y}{F_{n}(z)}.\label{3.22}
       \end{align}
 \end{lem3.1}
 \noindent
{\it Proof.}~Equation (\ref{3.19}) follows using the definition
 (\ref{3.7}) of $\phi$ as well as relations (\ref{2.15a})-(\ref{2.15c}).
Relations (\ref{3.20})-(\ref{3.22}) are clear from (\ref{2.19}), (\ref{3.18}), and (\ref{3.7}).
\quad $\square$

\vspace{0.1cm}
 The properties of $\psi(P,x,x_0)$ are summarized in the following lemma.
\newtheorem{lem3.2}[lem3.1]{Lemma}
 \begin{lem3.2}\label{lemma3.2}
    Suppose $(\ref{2.1})$, assume the $n$th stationary $HS2$ equation $(\ref{2.34})$ holds, and
    let
    $P=(z,y)\in \mathcal{K}_{n}\setminus \{P_{\infty_+}, P_{\infty_-},P_0\},$
    $(x,x_0) \in \mathbb{R}^2$. Then $\psi_1(P,x,x_0), \psi_2(P,x,x_0)$ satisfy
      \begin{align}
       & \psi_1(P,x,x_0)=\Big(\frac{F_{n}(z,x)}{F_{n}(z,x_0)}\Big)^{1/2}
         \mathrm{exp}\left(\frac{y}{z}
          \int_{x_0}^x F_{n}(z,x^\prime)^{-1} dx^\prime
          \right), \label{3.26} \\
       &
          \psi_1(P,x,x_0)\psi_1(P^\ast,x,x_0)=\frac{F_{n}(z,x)}{F_{n}(z,x_0)}, \label{3.27}
          \\
       &
           \psi_2(P,x,x_0)\psi_2(P^\ast,x,x_0)=-\frac{H_n(z,x)}{z^2 F_{n}(z,x_0)}, \label{3.28}
           \\
       &
           \psi_1(P,x,x_0)\psi_2(P^\ast,x,x_0)
        +\psi_1(P^\ast,x,x_0)\psi_2(P,x,x_0)
        =2\frac{G_n(z,x)}{F_{n}(z,x_0)}, \label{3.29}
        \\
       &
         \psi_1(P,x,x_0)\psi_2(P^\ast,x,x_0)
        -\psi_1(P^\ast,x,x_0)\psi_2(P,x,x_0)
        =\frac{-2y}{z F_{n}(z,x_0)}. \label{3.30}
      \end{align}
 \end{lem3.2}
\noindent
{\it Proof.}~Equation (\ref{3.26}) is a consequence of (\ref{2.15a}), (\ref{3.6}), and (\ref{3.7}).
Equation (\ref{3.27}) is clear from (\ref{3.26}) and (\ref{3.28}) is a consequence of (\ref{3.8}),
(\ref{3.20}), and (\ref{3.27}). Equation (\ref{3.29}) follows using (\ref{3.8}), (\ref{3.21}), and
(\ref{3.27}). Finally, (\ref{3.30}) follows from (\ref{3.8}), (\ref{3.22}), and (\ref{3.27}).
\quad $\square$
\vspace{0.1cm}

In Lemma \ref{lemma3.2}, we denote by
      $$\psi_1(P)=\psi_{1,+}, ~ \psi_1(P^\ast)=\psi_{1,-},~
       \psi_2(P)=\psi_{2,+}, ~ \psi_2(P^\ast)=\psi_{2,-},$$
and then (\ref{3.27})-(\ref{3.30}) imply
     \begin{equation}\label{3.36}
        (\psi_{1,+}\psi_{2,-}-\psi_{1,-}\psi_{2,+})^2=
          (\psi_{1,+}\psi_{2,-}+\psi_{1,-}\psi_{2,+})^2
          -4\psi_{1,+}\psi_{2,-}\psi_{1,-}\psi_{2,+},
     \end{equation}
which is equivalent to the basic identity (\ref{2.19}),
$z^2G_n^2+F_{n}H_n=R_{2n+2}$. This fact reveals the relations
between our approach and the algebro-geometric solutions of the HS2
hierarchy.

\newtheorem{rem3.3}[lem3.1]{Remark}
  \begin{rem3.3}
  The Baker-Akhiezer function $\psi$ of the stationary $HS2$ hierarchy
    is formally analogous to that defined
      in the context of KdV or AKNS
    hierarchies. However, its actual properties in a neighborhood
     of its essential singularity will feature characteristic differences
     to standard Baker-Akhiezer functions
     (cf. Remark $\ref{remark4.2}$).
  \end{rem3.3}

Next, we derive Dubrovin-type equations, that is,
first-order coupled systems of differential equations that govern the
dynamics of $\mu_j(x)$ and $\nu_l(x)$
with respect to variations of $x$.

\newtheorem{lem3.4}[lem3.1]{Lemma}
    \begin{lem3.4}\label{lemma3.4}
      Assume $(\ref{2.1})$ and the $n$th stationary $HS2$ equation
   $(\ref{2.34})$ holds subject to the constraint $(\ref{2.26a})$.
   \begin{itemize}
      \item[\emph{(i)}]
     Suppose that the zeros $\{\mu_j(x)\}_{j=0,\ldots,n}$
     of $F_{n}(z,x)$ remain distinct for $ x \in
     \Omega_\mu,$ where $\Omega_\mu \subseteq \mathbb{R}$ is an open
     interval, then
     $\{\mu_j(x)\}_{j=0,\ldots,n}$ satisfy the system of
     differential equations,
    \begin{equation}\label{3.38}
           \mu_{j,x}= 4 \frac{ y(\hat{\mu}_j)}{\mu_j}
            \prod_{\scriptstyle k=0 \atop \scriptstyle k \neq j }^{n}
            (\mu_j(x)-\mu_k(x))^{-1}, \quad j=0,\ldots,n,
        \end{equation}
   with initial conditions
       \begin{equation}\label{3.39}
         \{\hat{\mu}_j(x_0)\}_{j=0,\ldots,n}
         \in \mathcal{K}_{n},
       \end{equation}
   for some fixed $x_0 \in \Omega_\mu$. The initial value
   problem $(\ref{3.38})$, $(\ref{3.39})$ has a unique solution
   satisfying
        \begin{equation}\label{3.40}
         \hat{\mu}_j \in C^\infty(\Omega_\mu,\mathcal{K}_{n}),
         \quad j=0,\ldots,n.
        \end{equation}
    \item[\emph{(ii)}]
    Suppose that the zeros $\{\nu_l(x)\}_{l=0,\ldots,n}$
     of $H_n(z,x)$ remain distinct for $ x \in
     \Omega_\nu,$ where $\Omega_\nu \subseteq \mathbb{R}$ is an open
     interval, then
     $\{\nu_l(x)\}_{l=0,\ldots,n}$ satisfy the system of
     differential equations,
    \begin{equation}\label{3.41}
           \nu_{l,x}=-2\frac{(\rho^2+u_{xx}\nu_l)y(\hat{\nu}_l)}{h_0~\nu_l}
            \prod_{\scriptstyle k=0 \atop \scriptstyle k \neq l }^{n}
            (\nu_l(x)-\nu_k(x))^{-1}, \quad l=0,\ldots,n,
        \end{equation}
   with initial conditions
       \begin{equation}\label{3.42}
         \{\hat{\nu}_l(x_0)\}_{l=0,\ldots,n}
         \in \mathcal{K}_{n},
       \end{equation}
   for some fixed $x_0 \in \Omega_\nu$. The initial value
   problem $(\ref{3.41})$, $(\ref{3.42})$ has a unique solution
   satisfying
        \begin{equation}\label{3.43}
         \hat{\nu}_l \in C^\infty(\Omega_\nu,\mathcal{K}_{n}),
         \quad l=0,\ldots,n.
        \end{equation}
   \end{itemize}
  \end{lem3.4}
\noindent
{\it Proof.}~It suffices to prove (\ref{3.38}) and (\ref{3.40})
 since the proof of (\ref{3.41}) and (\ref{3.43}) follow in
an identical manner. Differentiating (\ref{3.9}) with respect
to $x$ then yields
   \begin{equation}\label{3.44}
   F_{n,x}(\mu_j)= -\frac{1}{2}\mu_{j,x}
   \prod_{\scriptstyle k=0 \atop \scriptstyle k \neq j }^{n}
   (\mu_j(x)-\mu_k(x)).
   \end{equation}
On the other hand, taking into account equation (\ref{2.15a}), one finds
       \begin{equation}\label{3.45}
        F_{n,x}(\mu_j)=2 G_n(\mu_j)
        = 2 \frac{y(\hat{\mu}_j)}{-\mu_j}.
       \end{equation}
Then combining equation (\ref{3.44}) with (\ref{3.45}) leads to ({\ref{3.38}). The
proof of smoothness assertion (\ref{3.40}) is analogous to
the KdV case in \cite{15}. \quad $\square$
\vspace{0.1cm}

Next, we turn to the trace formulas of the HS2 invariants, that
is, expressions of $f_{l}$ and $h_{l}$ in terms of symmetric
functions of the zeros $\mu_j$ and $\nu_l$ of $F_{n}$ and $H_n$,
respectively. For simplicity, we just record the simplest case.
\newtheorem{lem3.5}[lem3.1]{Lemma}
 \begin{lem3.5}\label{lemma3.5}
  Suppose $(\ref{2.1})$, assume the $n$th stationary $HS2$ equation
     $(\ref{2.34})$ holds, and let $x\in\mathbb{R}.$ Then
    \begin{equation}\label{3.46}
    u(x)= \frac{1}{2}\sum_{j=0}^n \mu_j(x) -\frac{1}{2}\sum_{m=0}^{2n+1} E_m.
    \end{equation}
 \end{lem3.5}
 \noindent
{\it Proof.}~Equation (\ref{3.46}) follows by considering the coefficient of $z^{n}$
in $F_{n}$ in (\ref{2.28}) and (\ref{3.9}),
 which yields
      \begin{equation}\label{3.47}
        -u+c_1=-\frac{1}{2}\sum_{j=0}^n \mu_j.
      \end{equation}
The constant $c_1$ can be determined by  a long straightforward calculation
considering the coefficient of $z^{2n+1}$ in (\ref{2.19}), which results in
     \begin{equation}\label{3.48}
        c_1=-\frac{1}{2}\sum_{m=0}^{2n+1} E_m.
     \end{equation}

\section{Stationary algebro-geometric solutions of HS2 hierarchy}
 In this section, we obtain explicit Riemann theta function
 representations for the meromorphic function $\phi$, and especially,
 for the solutions $u, \rho$ of the stationary HS2 hierarchy.

We begin with the asymptotic properties of $\phi$ and
$\psi_j,j=1,2$.

\newtheorem{lem4.1}{Lemma}[section]
   \begin{lem4.1}\label{lemma4.1}
    Suppose $(\ref{2.1})$, assume the $n$th stationary $HS2$ equation $(\ref{2.34})$
    holds, and let
    $P=(z,y)
    \in \mathcal{K}_n \setminus \{P_{\infty_+}, P_{\infty_-}, P_0\},$ $(x,x_0) \in
    \mathbb{R}^2$. Then
     \begin{align}
       & \phi(P) \underset{\zeta \rightarrow 0}{=}
         -u_x(x)+O(\zeta), \quad P \rightarrow P_{\infty_\pm}, \quad \zeta=z^{-1},
                \label{4.1} \\
       &
       \phi(P) \underset{\zeta \rightarrow 0}{=}
         i \rho(x)+\frac{i u_{xx}(x)-\rho_x(x)}{2\rho(x)} \zeta
          +O(\zeta^2),
           \quad P \rightarrow P_{0}, \quad \zeta=z,\label{4.2}
     \end{align}
and
     \begin{align}
      & \psi_1(P,x,x_0) \underset{\zeta \rightarrow 0}{=}
       \mathrm{exp}\left((u(x_0)-u(x))\zeta+O(\zeta^2)\right),
        \quad
        P \rightarrow P_{\infty_\pm},\quad \zeta=z^{-1}, \label{4.3}
         \\
      &
      \psi_2(P,x,x_0) \underset{\zeta \rightarrow 0}{=}O(\zeta)~
        \mathrm{exp}\left((u(x_0)-u(x))\zeta+O(\zeta^2)\right),  \quad
P \rightarrow P_{\infty_\pm},\quad \zeta=z^{-1}, \label{4.4}
        \\
     &
      \psi_1(P,x,x_0) \underset{\zeta \rightarrow 0}{=}
         \mathrm{exp} \left( \frac{i}{\zeta} \int_{x_0}^x dx^\prime
          ~\rho(x^\prime)
          +O(1) \right),
          \quad
           P \rightarrow P_{0}, \quad \zeta=z, \label{4.5}
          \\
     &
      \psi_2(P,x,x_0) \underset{\zeta \rightarrow 0}{=}
         O(\zeta^{-1})~
        \mathrm{exp} \left( \frac{i}{\zeta} \int_{x_0}^x dx^\prime
          ~\rho(x^\prime)
          +O(1) \right), \quad
           P \rightarrow P_{0}, \quad \zeta=z.  \label{4.6}
     \end{align}
   \end{lem4.1}
\noindent
  {\it Proof.}~The existence of the
asymptotic expansions of $\phi$ in terms of the appropriate
local coordinates $\zeta=z^{-1}$
near $P_{\infty_\pm}$ and $\zeta=z$ near $P_0$ is
clear from its explicit
expression in (\ref{3.7}). Next, we compute the coefficients
of these expansions utilizing the Riccati-type equation
 (\ref{3.19}). Indeed,
inserting the ansatz
\begin{equation}\label{4.7}
     \phi \underset{z \rightarrow \infty}{=}
     \phi_0 +\phi_1 z^{-1}+O(z^{-2})
  \end{equation}
into (\ref{3.19}) and comparing the same powers of $z$ then yields
(\ref{4.1}). Similarly, inserting the ansatz
    \begin{equation}\label{4.8}
     \phi \underset{z \rightarrow 0}{=} \phi_0+
       \phi_{1} z +O(z^2)
    \end{equation}
into (\ref{3.19}) and comparing the same powers of $z$ then yields
(\ref{4.2}). Finally, expansions
(\ref{4.3})-(\ref{4.6}) follow from (\ref{3.6}), (\ref{3.8}), (\ref{4.1}), and
(\ref{4.2}). \quad $\square$

\newtheorem{rem4.2}[lem4.1]{Remark}
  \begin{rem4.2}\label{remark4.2}
      We note the unusual fact
    that $P_0$, as opposed to $P_{\infty\pm}$, is the essential singularity
    of $\psi_j$, $j=1,2$.
    In addition, one easily finds the
    leading-order exponential term in $\psi_j$, $j=1,2,$ near $P_0$ is
    $x$-dependent, which makes matters worse.
    This is in sharp contrast to standard
    Baker-Akhiezer functions that typically feature a linear
    behavior
    with respect to $x$ in
    connection with their essential singularities of the type
    $\mathrm{exp}(c(x-x_0)\zeta^{-1})$
    near $\zeta=0$.
      \end{rem4.2}

Next, we introduce the holomorphic differentials $\eta_l(P)$ on
$\mathcal{K}_{n}$
      \begin{equation}\label{4.10}
        \eta_l(P)= \frac{a~z^{l-1}}{y(P)} dz,
        \quad l=1,\ldots,n,
      \end{equation}
and choose a homology basis $\{a_j,b_j\}_{j=1}^{n}$ on
$\mathcal{K}_{n}$ in such a way that the intersection matrix of the
cycles satisfies
$$a_j \circ b_k =\delta_{j,k},\quad a_j \circ a_k=0, \quad
b_j \circ   b_k=0, \quad j,k=1,\ldots, n.$$
Associated with $\mathcal{K}_n$, one
introduces an invertible
matrix $E \in GL(n, \mathbb{C})$
   \begin{equation}\label{4.11}
          \begin{split}
        & E=(E_{j,k})_{n \times n}, \quad E_{j,k}=
           \int_{a_k} \eta_j, \\
        &  \underline{c}(k)=(c_1(k),\ldots, c_{n}(k)), \quad
           c_j(k)=(E^{-1})_{j,k},
           \end{split}
   \end{equation}
and the normalized holomorphic differentials
        \begin{equation}\label{4.12}
          \omega_j= \sum_{l=1}^{n} c_j(l)\eta_l, \quad
          \int_{a_k} \omega_j = \delta_{j,k}, \quad
          \int_{b_k} \omega_j= \tau_{j,k}, \quad
          j,k=1, \ldots ,n.
        \end{equation}
Apparently, the matrix $\tau$ is symmetric and has a
positive-definite imaginary part.

We choose a fixed base point $Q_0 \in \mathcal{K}_{n} \setminus
\{\hat{\mu}_0(x), \hat{\nu}_0(x)\}$.
The Abel
maps $\underline{A}_{Q_0}(\cdot) $ and
$\underline{\alpha}_{Q_0}(\cdot)$ are defined by
\begin{equation}\label{4.24}
    \begin{split}
    &
   \underline{A}_{Q_0}:\mathcal{K}_{n} \rightarrow
   J(\mathcal{K}_{n})=\mathbb{C}^{n}/L_{n}, \\
   &
   P \mapsto \underline{A}_{Q_0} (P)=(A_{Q_0,1}(P),\ldots,
  A_{Q_0,n} (P))
    =\left(\int_{Q_0}^P\omega_1,\ldots,\int_{Q_0}^P\omega_{n}\right)
  (\mathrm{mod}~L_{n})
  \end{split}
  \end{equation}
and
   \begin{equation}\label{4.25}
     \begin{split}
   & \underline{\alpha}_{Q_0}:
   \mathrm{Div}(\mathcal{K}_{n}) \rightarrow
   J(\mathcal{K}_{n}),\\
   & \mathcal{D} \mapsto \underline{\alpha}_{Q_0}
   (\mathcal{D})= \sum_{P\in \mathcal{K}_{n}}
    \mathcal{D}(P)\underline{A}_{Q_0} (P),
    \end{split}
    \end{equation}
where $L_{n}=\{\underline{z}\in \mathbb{C}^{n}|
           ~\underline{z}=\underline{N}+\tau\underline{M},
           ~\underline{N},~\underline{M}\in \mathbb{Z}^{n}\}.$

Let $\Phi_{n}^{(j)}(\bar{\mu}),$ $\Psi_{n+1}(\bar{\mu})$ be the symmetric functions of $\{\mu_j\}_{j=0,\ldots,n},$
 \begin{equation}\label{4.13}
        \Phi_{n}^{(j)}(\bar{\mu})=(-1)^{n}
          \prod_{\scriptstyle p=0 \atop \scriptstyle p \neq j}^n \mu_p,
    \quad
        \Psi_{n+1}(\bar{\mu})=(-1)^{n+1} \prod_{p=0}^n \mu_p.
     \end{equation}
     Here, $\bar{\mu}(x)=(\mu_0(x),\mu_1(x),\ldots,
   \mu_n(x))=\mu_0(x) \underline{\mu}(x).$
\vspace{0.15cm}

The following result shows the
nonlinearity of the Abel map with respect to
the variable $x$,
which
indicates a characteristic
difference between the HS2 hierarchy and other completely integrable
systems such as the KdV and AKNS hierarchies.

\newtheorem{the4.3}[lem4.1]{Theorem}
  \begin{the4.3}\label{theorem4.3}
    Assume $(\ref{2.26a})$ and
    suppose that
    $\{\hat{\mu}_j(x)\}_{j=0,\ldots,n}$ satisfies the stationary
    Dubrovin equations $(\ref{3.38})$ on an open  interval $\Omega_\mu\subseteq\mathbb{R}$ such that
    $\mu_j(x),j=0,\ldots,n,$
    remain distinct and nonzero for $ x \in
      \Omega_\mu.$ Introducing the associated divisor
     $\mathcal{D}_{ \hat{\mu}_0(x)\underline{\hat{\mu}}(x)}$, one computes
      \begin{equation}\label{4.15}
            \partial_x
            \underline{\alpha}_{Q_0}( \mathcal{D}_{ \hat{\mu}_0(x)\underline{\hat{\mu}}(x)})
            = -\frac{4a}{ \Psi_{n+1}(\bar{\mu}(x))}
             \underline{c}(1),
            \quad    x \in  \Omega_\mu.
        \end{equation}
   In particular, the Abel map does not linearize the divisor
   $\mathcal{D}_{\hat{\mu}_0(x)\underline{\hat{\mu}}(x)}$ on $ \Omega_\mu$.
  \end{the4.3}
\noindent
{\it Proof.}~Let $x\in \Omega_\mu.$ Then,
using
 \begin{equation}\label{4.16}
 \frac{1}{\mu_j}=
        \frac{\prod_{\scriptstyle p=0 \atop \scriptstyle p \neq j}^n \mu_p}
            {\prod_{p=0}^n \mu_p}
        =-\frac{
        \Phi_{n}^{(j)}(\bar{\mu})}{\Psi_{n+1}(\bar{\mu})},
        \quad j=0,\ldots,n,
        \end{equation}
 one obtains
\begin{align}\label{4.18}
      \partial_x  \underline{\alpha}_{Q_0}(\mathcal{D}_{\hat{\mu}_0(x)\underline{\hat{\mu}}(x)})
 &=
 \partial_x  \left(\sum_{j=0}^n \int_{Q_0}^{\hat{\mu}_j} \underline{\omega} \right)
 =\sum_{j=0}^n \mu_{j,x} \sum_{k=1}^n \underline{c}(k)
  \frac{a~\mu_j^{k-1}}{y(\hat{\mu}_j)}
      \nonumber \\
   &=
    \sum_{j=0}^n \sum_{k=1}^n
    \frac{4a~\mu_j^{k-1}}{\mu_j}
    \frac{1}
    {\prod_{\scriptstyle l=0 \atop \scriptstyle l \neq j }^{n}(\mu_j-\mu_l)}
    \underline{c}(k)
     \nonumber \\
    &=
     -\frac{4a}{\Psi_{n+1}(\bar{\mu})}
     \sum_{j=0}^n \sum_{k=1}^n \underline{c}(k)
       \frac{\mu_j^{k-1}}
    {\prod_{\scriptstyle l=0 \atop \scriptstyle l \neq j }^{n}(\mu_j-\mu_l)}
     \Phi_{n}^{(j)}(\bar{\mu})
       \nonumber \\
     &=
     -\frac{4a}{\Psi_{n+1}(\bar{\mu})}
     \sum_{j=0}^n \sum_{k=1}^n \underline{c}(k)
     (U_{n+1}(\bar{\mu}))_{k,j}(U_{n+1}(\bar{\mu}))_{j,1}^{-1}
       \nonumber \\
    &=
     -\frac{4a}{\Psi_{n+1}(\bar{\mu})}
      \sum_{k=1}^n \underline{c}(k) \delta_{k,1}
      \nonumber \\
    &=
      -\frac{4a}{\Psi_{n+1}(\bar{\mu})}
        \underline{c}(1),
    \end{align}
where we used the  notation
 $\underline{\omega}=(\omega_1,\ldots,\omega_n)$,
 and
the relations (cf.(E.13), (E.14) \cite{15}),
        \begin{equation}\label{4.19}
         U_{n+1}(\bar{\mu})=
       \left( \frac{\mu_j^{k-1}}
    {\prod_{\scriptstyle l=0 \atop \scriptstyle l \neq j }^{n}(\mu_j-\mu_l)}
    \right)_{\scriptstyle j=0 \atop \scriptstyle k=1}^n, \quad
    U_{n+1}(\bar{\mu})^{-1}=
     \left(\Phi_{n}^{(j)}(\bar{\mu})\right)_{\scriptstyle j=0 }^n.
    \end{equation}

The analogous results hold for the corresponding divisor
$\mathcal{D}_{\hat{\nu}_0(x)\underline{\hat{\nu}}(x)}$ associated with
$\phi(P,x).$

\vspace{0.1cm}

Next, we introduce
   \begin{equation}\label{4.20}
     \begin{split}
    & ~~\underline{\widehat{B}}_{Q_0}: \mathcal{K}_n \setminus
    \{P_{\infty_+}, P_{\infty_-}\} \rightarrow \mathbb{C}^n, \\
    & ~~~~~P \mapsto \underline{\widehat{B}}_{Q_0}(P)
    =(\widehat{B}_{Q_0,1},\ldots,\widehat{B}_{Q_0,n}) \\
    & ~~~~~~~~~~~~~~~~~
     =
      \left\{
        \begin{array}{ll}
          \int_{Q_0}^P \tilde{\omega}_{P_{\infty_+}, P_{\infty_-}}^{(3)}, & \hbox{$n=1$,} \\
           \Big(\int_{Q_0}^P \eta_2,\ldots, \int_{Q_0}^P \eta_n,
           \int_{Q_0}^P \tilde{\omega}_{P_{\infty_+}, P_{\infty_-}}^{(3)}
          \Big), & \hbox{$n \geq 2$,}
        \end{array}
      \right.
     \end{split}
   \end{equation}
where
$\tilde{\omega}_{P_{\infty_+}, P_{\infty_-}}^{(3)}=
a~z^{n}dz/y(P)$ (cf.(F.53) \cite{15})
and
\begin{equation}\label{4.21}
     \begin{split}
       & \underline{\hat{\beta}}_{Q_0}:
         \mathrm{Sym}^n(
       \mathcal{K}_n \setminus
    \{P_{\infty_+}, P_{\infty_-}\}) \rightarrow \mathbb{C}^n, \\
      & \mathcal{D}_{\underline{Q}} \mapsto
      \underline{\hat{\beta}}_{Q_0}(\mathcal{D}_{\underline{Q}})
      =\sum_{j=1}^n \underline{\widehat{B}}_{Q_0}(Q_j),
      \quad
      \underline{Q}=\{Q_1,\ldots,Q_n\}
      \in \mathrm{Sym}^n( \mathcal{K}_n \setminus
    \{P_{\infty_+}, P_{\infty_-}\}),
     \end{split}
   \end{equation}
choosing identical paths of integration from $Q_0$ to $P$
in all integrals in (\ref{4.20}) and (\ref{4.21}).
Then, one obtains the following result.

\newtheorem{the4.4}[lem4.1]{Corollary}
  \begin{the4.4}\label{corollary4.4}
    Assume $(\ref{2.26a})$ and
    suppose that
    $\{\hat{\mu}_j(x)\}_{j=0,\ldots,n}$ satisfies the stationary
    Dubrovin equations $(\ref{3.38})$ on an open interval $\Omega_\mu\subseteq\mathbb{R}$ such that
    $\mu_j(x),j=0,\ldots,n,$
    remain distinct and nonzero for $ x \in
    \Omega_\mu.$ Then one computes
    \begin{align}
       & \partial_x \sum_{j=0}^n \int_{Q_0}^{\hat{\mu}_j(x)} \eta_1
        = -\frac{4a}{\Psi_{n+1}(\bar{\mu}(x))},
        \quad x\in \Omega_\mu, \label{4.22}
        \\
       &
       \partial_x \underline{\hat{\beta}}
        (\mathcal{D}_{ \underline{\hat{\mu}}(x)})
        =
        \left\{
          \begin{array}{ll}
            4a, & \hbox{$n=1$,} \\
            4a~(0,\ldots,0,1), & \hbox{$n\geq 2$,}
          \end{array}
        \right.
        \quad  x\in \Omega_\mu.\label{4.23}
    \end{align}
  \end{the4.4}
\noindent
{\it Proof.}~Equation (\ref{4.22}) is a special case
(\ref{4.15}). Equation (\ref{4.23}) follows from (\ref{4.18}).
  \quad  $\square$
\vspace{0.1cm}

The fact
that the Abel map
does not
provide the proper change of variables
to linearize the divisor
  $\mathcal{D}_{\hat{\mu}_0(x) \underline{\hat{\mu}}(x)}$
in the HS2 context is in sharp contrast to
 standard integrable soliton equations such as the KdV and AKNS hierarchies. However, the change of variables
     \begin{equation}\label{4.33}
        x \mapsto \tilde{x}= \int^x dx^\prime
        \Big( \frac{4a}{\Psi_{n+1}(\bar{\mu}(x^\prime))}
                \Big)
     \end{equation}
linearizes the Abel map
$\underline{A}_{Q_0}(\mathcal{D}_{\hat{\tilde{\mu}}_0(\tilde{x})\underline{\hat{\tilde{\mu}}}(\tilde{x})}),$
$\tilde{\mu}_j(\tilde{x})=\mu_j(x),j=0,\ldots,n.$ The intricate
relation between the variable $x$ and $\tilde{x}$  is
detailed in (\ref{4.36}).

\vspace{0.1cm}

Next, given the Riemann surface $\mathcal{K}_n$ and the homology basis $\{a_j,b_j\}_{j=1,\ldots,n}$,
one introduces the Riemann theta function by
$$\theta(\underline{z})=\sum_{\underline{n}\in\mathbb{Z}^n}\exp\Big(2\pi i(\underline{n},
\underline{z})+\pi i(\underline{n},\tau\underline{n})\Big),~~\underline{z}\in\mathbb{C}^n,$$
where $(\underline{A},\underline{B})=\sum_{j=1}^{n}\overline{A}_jB_j$ denotes the scalar product in $\mathbb{C}^n.$

Let
   \begin{equation}\label{4.26}
    \omega_{\hat{\mu}_0(x), \hat{\nu}_0(x) }^{(3)}(P)=
     \frac{a}{y} \prod_{j=1}^n(z-\lambda_j) dz
    \end{equation}
be the normalized differential of the third kind  holomorphic on
$\mathcal{K}_{n} \setminus \{\hat{\mu}_0(x), \hat{\nu}_0(x)\}$ with simple poles at
$\hat{\mu}_0(x)$ and $\hat{\nu}_0(x)$  and residues $1$ and $-1$, respectively,
 \begin{align}
 & \omega_{\hat{\mu}_0(x),\hat{\nu}_0(x) }^{(3)}(P) \underset
              {\zeta \rightarrow 0}{=} (\zeta^{-1}+O(1))d \zeta,
              \quad \textrm{as $P \rightarrow \hat{\mu}_0(x),$}\label{4.27b}\\
             & \omega_{\hat{\mu}_0(x), \hat{\nu}_0(x) }^{(3)}(P) \underset
              {\zeta \rightarrow 0}{=} (-\zeta^{-1}+O(1))d \zeta,
              \quad \textrm{as $P \rightarrow \hat{\nu}_0(x),$}\label{4.27a}
              \end{align}
where $\zeta$ in (\ref{4.27b}) and (\ref{4.27a}) denotes the corresponding local coordinate
near $\hat{\mu}_0(x)$ and $ \hat{\nu}_0(x)$ (cf. Sect.3, also Appendix C \cite{15}).
The constants $\{\lambda_j\}_{j=1,\ldots,n}$ in (\ref{4.26}) are determined by
the normalization condition
      $$\int_{a_k} \omega_{\hat{\mu}_0(x), \hat{\nu}_0(x) }^{(3)}=0,
      \quad k=1,\ldots,n.$$
Then
 \begin{equation} \label{4.29}
   \int_{Q_0}^P \omega_{\hat{\mu}_0(x),\hat{\nu}_0(x) }^{(3)}(P) \underset
  {\zeta \rightarrow 0}{=} \mathrm{ln} \zeta +
  e_0+O(\zeta),
  \quad \textrm{as $P \rightarrow \hat{\mu}_0(x),$ }
  \end{equation}
  \begin{equation}\label{4.30}
    \int_{Q_0}^P \omega_{\hat{\mu}_0(x), \hat{\nu}_0(x) }^{(3)}(P) \underset
    {\zeta \rightarrow 0}{=} -\mathrm{ln} \zeta +
    d_0+O(\zeta),
     \quad \textrm{as $P \rightarrow \hat{\nu}_0(x),$}
    \end{equation}
for some constants $e_0,d_0 \in \mathbb{C}$.
We also record
  \begin{equation}\label{4.31}
   \underline{A}_{Q_0}(P)-\underline{A}_{Q_0}(P_{\infty_\pm})
   \underset {\zeta \rightarrow 0}{=}
   \pm \underline{U}\zeta+O(\zeta^2),
   \quad \textrm{as $P \rightarrow P_{\infty_\pm},$}
   \quad \underline{U}=\underline{c}(n).
  \end{equation}
In the following, it will be convenient to introduce
the abbreviations
    \begin{eqnarray}\label{4.32}
         &&
           \underline{z}(P,\underline{Q})= \underline{\Xi}_{Q_0}
           -\underline{A}_{Q_0}(P)+\underline{\alpha}_{Q_0}
             (\mathcal{D}_{\underline{Q}}), \nonumber \\
          &&
           P\in \mathcal{K}_{n},\,
          \underline{Q}=(Q_1,\ldots,Q_{n})\in
          \mathrm{Sym}^{n}(\mathcal{K}_{n}),
         \end{eqnarray}
where $\underline{\Xi}_{Q_0}$ is
the vector of Riemann constants (cf.(A.45) \cite{15}).
It turns out that $\underline{z}(\cdot,\underline{Q}) $ is independent of the
choice of base point $Q_0$\,(cf.(A.52),\,(A.53) \cite{15}).

Based on above preparations, we will give
explicit representations for the meromorphic function $\phi$ and the
stationary HS2 solutions $u, \rho$ in terms of the Riemann theta function
associated with $\mathcal{K}_{n}$.

\newtheorem{the4.5}[lem4.1]{Theorem}
 \begin{the4.5}\label{theorem4.5}
 Suppose
 $(\ref{2.1})$, and assume the
 $n$th stationary $HS2$ equation
  $(\ref{2.34})$ holds on $ \Omega$ subject to the
  constraint $(\ref{2.26a})$. Moreover, let
$P=(z,y) \in \mathcal{K}_n \setminus \{P_0\}$ and $x \in  \Omega$,
where $ \Omega \subseteq \mathbb{R}$ is an open interval. In
addition, suppose that $\mathcal{D}_{\underline{\hat{\mu}}(x)}$, or
equivalently, $\mathcal{D}_{\underline{\hat{\nu}}(x)}$ is nonspecial
for $x\in \Omega$. Then, $\phi$, $u,$ and $\rho$ admit the following
representations
\begin{align}
  &  \phi(P,x)= i \rho(x)
    \frac{\theta(\underline{z}(P,\underline{\hat{\nu}}(x)))
            \theta(\underline{z}(P_0,\underline{\hat{\mu}}(x)))}
            {\theta(\underline{z}(P_0,\underline{\hat{\nu}}(x)))
            \theta(\underline{z}(P,\underline{\hat{\mu}}(x)))}
                   \mathrm{exp}\left(e_0
            -\int_{Q_0}^P \omega_{\hat{\mu}_0(x), \hat{\nu}_0(x)}^{(3)}\right),
             \label{4.34}
       \\
  &
  u(x)=-\frac{1}{2} \sum_{m=0}^{2n+1} E_m +\frac{1}{2} \sum_{j=1}^n \lambda_j
         -\frac{1}{2} \sum_{j=1}^nU_j \partial_{\omega_j} \mathrm{ln}
\left(\frac{\theta(\underline{z}(P_{\infty_+},\underline{\hat{\mu}}(x))+\underline{\omega})}
 {\theta(\underline{z}(P_{\infty_-},\underline{\hat{\mu}}(x))+\underline{\omega})}\right)
     \Big|_{\underline{\omega}=0}, \label{4.35} \\
  &
   \rho(x)=i u_x (x) \frac{\theta(\underline{z}(P_0,\underline{\hat{\nu}}(x)))
            \theta(\underline{z}(P_{\infty_+},\underline{\hat{\mu}}(x)))}
            {\theta(\underline{z}(P_{\infty_+},\underline{\hat{\nu}}(x)))
            \theta(\underline{z}(P_0,\underline{\hat{\mu}}(x)))}.\label{4.35a1}
  \end{align}
Moreover, let $\widetilde{\Omega}\subseteq\Omega$
be such that
$\mu_j,$ $j=0,\ldots,n,$ are nonvanishing on $\widetilde{\Omega}.$
Then, the constraint
  \begin{align}\label{4.36}
        4a(x-x_0)
        &=-4a \int_{x_0}^x
        \frac{dx^\prime}{\prod_{k=0}^n \mu_k(x^\prime)}
        \sum_{j=1}^n
        \Big(\int_{a_j} \tilde{\omega}_{P_{\infty_+} P_{\infty_-}}^{(3)} \Big)
        c_j(1) \nonumber \\
        &+
         \mathrm{ln}
        \left(
        \frac{\theta(\underline{z}(P_{\infty_-},\underline{\hat{\mu}}(x_0)))
            \theta(\underline{z}(P_{\infty_+},\underline{\hat{\mu}}(x)))}
            {\theta(\underline{z}(P_{\infty_+},\underline{\hat{\mu}}(x_0)))
            \theta(\underline{z}(P_{\infty_-},\underline{\hat{\mu}}(x)))}
        \right),\ \ x,x_0\in\widetilde{\Omega}
  \end{align}
holds, with
\begin{align}\label{4.37}
      \underline{\hat{\alpha}}_{Q_0}( \mathcal{D}_{ \hat{\mu}_0(x)\underline{\hat{\mu}}(x)})
      &=
      \underline{\hat{\alpha}}_{Q_0}(\mathcal{D}_{ \hat{\mu}_0(x_0)\underline{\hat{\mu}}(x_0)})
       -4a \int_{x_0}^x
       \frac{dx^\prime}{ \Psi_{n+1}(\bar{\mu}(x^\prime))}
             \underline{c}(1)
          \nonumber \\
       &=
       \underline{\hat{\alpha}}_{Q_0}(\mathcal{D}_{\hat{\mu}_0(x) \underline{\hat{\mu}}(x_0)})
       - \underline{c}(1)(\tilde{x}-\tilde{x}_0),\ \ x\in\widetilde{\Omega}.
     \end{align}
\end{the4.5}
\noindent
{\it Proof.}~First, we temporarily assume that
    \begin{equation}\label{4.38}
      \mu_j(x)\neq \mu_{j^\prime}(x), \quad \nu_k(x)\neq \nu_{k^\prime}(x)
      \quad \textrm{for $j\neq j^\prime, k\neq k^\prime$ and
      $x\in\widetilde{\Omega}$},
      \end{equation}
 for appropriate $\widetilde{\Omega}\subseteq\Omega$. Since by (\ref{3.14}), $\mathcal
 {D}_{\hat{\nu}_0  \underline{\hat{\nu}}}\sim
 \mathcal {D}_{\hat{\mu}_0  \underline{\hat{\mu}}}$, and
 $(\hat{\mu}_0) ^\ast \notin\{\hat{\mu}_1,\ldots,\hat{\mu}_n \}$
 by hypothesis, one can use Theorem A.31 \cite{15} to
 conclude that $\mathcal {D}_{\underline{\hat{\nu}}}
 \in \textrm{Sym}^n(\mathcal {K}_n)$ is nonspecial. This
 argument is of course symmetric with respect to
 $\underline{\hat{\mu}}$ and $\underline{\hat{\nu}}$. Thus, $\mathcal
 {D}_{\underline{\hat{\mu}}}$ is nonspecial if and only
 if $\mathcal{D}_{\underline{\hat{\nu}}}$ is.

Next, we derive the representations of $\phi$, $u$, and $\rho$  in terms of the Riemann theta
function. A special case of Riemann's vanishing theorem (cf.\,Theorem A.26\,\cite{15})
yields
  \begin{equation}\label{4.39}
        \theta(\underline{\Xi}_{Q_0}-\underline{A}_{Q_0}(P)+\underline{\alpha}_{Q_0}(\mathcal
        {D}_{\underline{Q}}))=0 \quad \textrm{ if and only if $P\in
        \{Q_1,\ldots,Q_n\}$}.
  \end{equation}
Therefore, the divisor (\ref{3.14}) shows that $\phi(P,x)$ has expression of the
type
\begin{equation}\label{4.40}
C(x)\frac{
\theta(\underline{\Xi}_{Q_0}-\underline{A}_{Q_0}(P)+\underline{\alpha}_{Q_0}(\mathcal
{D}_{\underline{\hat{\nu}}(x)}))}{
\theta(\underline{\Xi}_{Q_0}-\underline{A}_{Q_0}(P)+\underline{\alpha}_{Q_0}(\mathcal
{D}_{\underline{\hat{\mu}}(x)}))} \mathrm{exp}\Big(e_0-\int_{Q_0}^P
\omega_{\hat{\mu}_0(x),\hat{\nu}_0(x) }^{(3)}
\Big),
\end{equation}
where $C(x)$ is independent of $P\in\mathcal {K}_n$. Then
 taking  into account
 the asymptotic expansion of $\phi(P,x)$ near $P_0$ in
(\ref{4.2}), we obtain (\ref{4.34}). The representation
(\ref{4.35}) for $u$ on $ \widetilde{\Omega}$ follows from trace
formula (\ref{3.46}) and the expression (F.59)\cite{15} for
$\sum_{j=0}^n \mu_j$. The representation (\ref{4.35a1}) for $\rho$ on $ \widetilde{\Omega}$
is clear from (\ref{4.1}) and (\ref{4.34}).
By continuity, (\ref{4.34}), (\ref{4.35}),  and
(\ref{4.35a1}) extend from $\widetilde{\Omega}$ to $\Omega$.
Assuming $\mu_j \neq 0$, $j=0,\ldots,n$,
the constraint (\ref{4.36}) follows by combining
(\ref{4.22}), (\ref{4.23}), and (F.58) \cite{15}.
Equation (\ref{4.37}) is clear from
(\ref{4.15}). Finally, (\ref{4.36}) and (\ref{4.37}) extend to
$\widetilde{\Omega}$ by continuity.
\quad $\square$

\newtheorem{rem4.6}[lem4.1]{Remark}
  \begin{rem4.6}
  We note that the stationary $HS2$ solutions $u, \rho$ in $(\ref{4.35})$ and $(\ref{4.35a1})$ are meromorphic
  quasi-periodic functions with respect to the new variable
  $\tilde{x}$ in $(\ref{4.33})$. In addition, the Abel map in $(\ref{4.37})$
  linearizes the divisor
  $\mathcal{D}_{\hat{\mu}_0(x) \underline{\hat{\mu}}(x)}$ with respect
  to $\tilde{x}$ under the constraint
  $(\ref{4.36})$.
  \end{rem4.6}

\newtheorem{rem4.8}[lem4.1]{Remark}
  \begin{rem4.8}\label{remark4.8}
    Since by $(\ref{3.14})$ $\mathcal
 {D}_{ \hat{\nu}_0  \underline{\hat{\nu}}}$ and
 $\mathcal {D}_{ \hat{\mu}_0 \underline{\hat{\mu}}}$
 are linearly equivalent, that is
    \begin{equation}\label{4.42}
    \underline{A}_{Q_0}(\hat{\mu}_0(x)) + \underline{\alpha}_{Q_0}
    (\mathcal{D}_{\underline{\hat{\mu}}(x)})
    =\underline{A}_{Q_0}(\hat{\nu}_0(x))+ \underline{\alpha}_{Q_0}
    (\mathcal{D}_{\underline{\hat{\nu}}(x)}).
    \end{equation}
Then one infers
    \begin{equation}\label{4.43}
        \underline{\alpha}_{Q_0}
    (\mathcal{D}_{\underline{\hat{\nu}}(x)})
    =\underline{\Delta}+ \underline{\alpha}_{Q_0}
    (\mathcal{D}_{\underline{\hat{\mu}}(x)}),
    \quad \underline{\Delta}=\underline{A}_{\hat{\nu}_0(x)}(\hat{\mu}_0(x)).
    \end{equation}
Hence, one can eliminate $\mathcal{D}_{\underline{\hat{\nu}}(x)}$
in $(\ref{4.34})$, in terms of
$\mathcal{D}_{\underline{\hat{\mu}}(x)}$ using
   \begin{equation}\label{4.44}
    \underline{z}(P,\underline{\hat{\nu}})=
    \underline{z}(P,\underline{\hat{\mu}})+\underline{\Delta},
    \quad P\in \mathcal{K}_n.
   \end{equation}
  \end{rem4.8}

\newtheorem{rem4.9}[lem4.1]{Remark}
  \begin{rem4.9}\label{remark4.9}
   We emphasized in Remark $\ref{remark4.2}$ that $\psi$ in $(\ref{3.6})$ and $(\ref{3.8})$
   differs from standard Baker-Akhiezer functions.
   Hence, one can not expect the usual theta function representation
   of $\psi_j$, $j=1,2,$ in terms of ratios of theta functions
   times an exponential term containing a
   meromorphic differential with a pole at the essential singularity
   of $\psi_j$ multiplied by $(x-x_0)$. However, using
   $(E.3)$ and $(F.59)~\cite{15}$, one computes
   \begin{align}\label{4.45}
   F_{n}(z)&=\frac{1}{2}z^{n+1}+\frac{1}{2}\sum_{k=0}^{n} \Psi_{n+1-k}(\bar{\mu}(x)) z^k
    \nonumber \\
    &=
    \frac{1}{2}z^{n+1}+\frac{1}{2}\sum_{k=1}^n \Big( \Psi_{n+1-k}(\underline{\lambda})
    -\sum_{j=1}^nc_j(k) \partial_{\omega_j}
    \mathrm{ln}
    \left(
    \frac{\theta(\underline{z}(P_{\infty_+},\underline{\hat{\mu}}(x))+\underline{\omega})}
    {\theta(\underline{z}(P_{\infty_-},\underline{\hat{\mu}}(x))+\underline{\omega})}
    \right)
    \Big|_{\underline{\omega}=0} \Big) z^{k}
       \nonumber \\
    &=
      \frac{1}{2}\prod_{j=0}^n(z-\lambda_j)
       -\frac{1}{2}\sum_{j=1}^n \sum_{k=1}^n
       c_j(k) \partial_{\omega_j}
    \mathrm{ln}
    \left(
    \frac{\theta(\underline{z}(P_{\infty_+},\underline{\hat{\mu}}(x))+\underline{\omega})}
    {\theta(\underline{z}(P_{\infty_-},\underline{\hat{\mu}}(x))+\underline{\omega})}
    \right)
    \Big|_{\underline{\omega}=0} z^{k},
   \end{align}
and  hence obtains the
theta function representation of $\psi_1$ upon inserting $(\ref{4.45})$ into $(\ref{3.26})$.
Then, the corresponding
theta function representation of $\psi_2$ follows by $(\ref{3.8})$
and $(\ref{4.34})$.
  \end{rem4.9}

At the end of this section, we turn to the initial value problem for
the stationary HS2 hierarchy. We will show that the solvability of the Dubrovin
equations (\ref{3.38}) on $\Omega_\mu \subseteq \mathbb{R}$ in fact
implies the stationary HS2 equation (\ref{2.34}) on $\Omega_\mu$.

\newtheorem{the4.10}[lem4.1]{Theorem}
 \begin{the4.10}\label{theorem4.10}
    Fix $n \in \mathbb{N}_0$,
    assume $(\ref{2.26a})$, and suppose that
     $\{\hat{\mu}_j\}_{j=0,\ldots,n}$ satisfies the stationary
    Dubrovin equations $(\ref{3.38})$ on $ \Omega_\mu$
    such that $\mu_j$, $j=0,\ldots,n$,
    remain distinct and nonzero on $
     \Omega_\mu,$ where $\Omega_\mu \subseteq \mathbb{R}$ is an open
     interval. Then, $u, \rho \in C^\infty(\Omega_\mu)$, defined by
      \begin{equation}\label{4.46}
        u=-\frac{1}{2} \sum_{m=0}^{2n+1} E_m + \frac{1}{2} \sum_{j=0}^n \mu_j,
      \end{equation}
and
     \begin{equation}
     \rho^2=u_x^2+2uu_{xx}-\frac{d^2}{dx^2}
     \left(
     \Psi_2(\bar{\mu})-u\sum_{m=0}^{2n+1} E_m
     \right),
     \end{equation}
     satisfy the $n$th stationary $HS2$ equation
   $(\ref{2.34}),$ that is,
     \begin{equation}\label{4.47}
        \mathrm{s}\textrm{-}\mathrm{HS2}_n(u,\rho)=0
        ~ \textrm{on $\Omega_\mu.$}
     \end{equation}
  \end{the4.10}
\noindent
{\it Proof.}~Given the solutions
   $\hat{\mu}_j=(\mu_j,y(\hat{\mu}_j))\in
   C^\infty(\Omega_\mu,\mathcal {K}_n),j=0,\ldots,n$ of (\ref{3.38}),
   we introduce
   \begin{eqnarray}\label{4.48}
     \begin{split}
       &F_{n}(z)=\frac{1}{2} \prod_{j=0}^n(z-\mu_j), \\
     &G_n(z)=\frac{1}{2}F_{n,x}(z),
     \end{split}
   \end{eqnarray}
    \textrm{on $\mathbb{C}\times\Omega_\mu$}.
 Taking into account (\ref{4.48}),
 the Dubrovin equations (\ref{3.38}) imply
     \begin{equation}\label{4.50}
       y(\hat{\mu}_j)= \frac{1}{4} \mu_j \mu_{j,x}
         \prod_{ \scriptstyle k=0 \atop \scriptstyle k \neq j}^n
         (\mu_j-\mu_k)
       = -\frac{1}{2} \mu_j F_{n,x}(\mu_j)
       =-\mu_j G_n(\mu_j).
      \end{equation}
  Hence
     \begin{equation}\label{4.51}
        R_{2n+2}(\mu_j)^2-\mu_j^2 G_n(\mu_j)^2=
        y(\hat{\mu}_j)^2-\mu_j^2 G_n(\mu_j)^2=0,
        \quad j=0,\ldots,n.
     \end{equation}
Next, we define a polynomial $H_n$ on $\mathbb{C}\times\Omega_\mu$
such that
    \begin{equation}\label{4.52}
      R_{2n+2}(z)-z^2G_n(z)^2=F_{n}(z)H_n(z).
    \end{equation}
 Such a polynomial $H_n$  exists since the left-hand side of (\ref{4.52})
 vanishes at $z=\mu_j,~j=0,\cdots,n$, by (\ref{4.51}). To
 determine the degree of $H_n$, using (\ref{4.48}), one computes
   \begin{equation}\label{4.53}
     R_{2n+2}(z)-z^2G_n(z)^2 \underset{|z| \rightarrow \infty}{=}
    \frac{1}{2}h_0 z^{2n+2}+O(z^{2n+1}).
   \end{equation}
Then combining (\ref{4.48}), (\ref{4.52}), and (\ref{4.53}),
  one infers that $H_n$ has degree $n+1$ with respect to
 $z$. Hence,
  we may write
      \begin{equation}\label{4.54}
        H_n(z)=h_0 \prod_{l=0}^{n}
        (z-\nu_l),
         \quad
         \textrm{on $\mathbb{C}\times\Omega_\mu$}.
        \end{equation}
Next, one defines the polynomial $P_{n}$ by
    \begin{equation}\label{4.55}
        P_{n}(z)=H_n(z)+(\rho^2+u_{xx}z)F_{n}(z)
        +z^2G_{n,x}(z).
    \end{equation}
 Using (\ref{4.48}) and (\ref{4.54}), one  infers that indeed
 $P_{n}$ has degree at most $n$. Differentiating
 (\ref{4.52}) with respect to $x$ yields
     \begin{equation}\label{4.56}
        2z^2G_n(z)G_{n,x}(z)+F_{n,x}(z)H_n(z)+F_{n}(z)H_{n,x}(z)=0.
    \end{equation}
Then
 multiplying (\ref{4.55}) by $G_n$ and replacing the term $G_nG_{n,x}$ with (\ref{4.56}) leads to
    \begin{align}\label{4.57}
     G_n(z)P_{n}(z)= F_{n}(z)((\rho^2+u_{xx}z)G_n(z)-\frac{1}{2}H_{n,x}(z))
       +
     (G_n(z)-\frac{1}{2}F_{n,x}(z)) H_n(z),
  \end{align}
and hence
    \begin{equation}\label{4.58}
        G_n(\mu_j)P_{n}(\mu_j)=0,
        \quad j=0,\ldots,n
    \end{equation}
 on $\Omega_\mu$. Restricting
    $x \in \Omega_\mu$ temporarily to  $x\in\widetilde{\Omega}_\mu $,
where
      \begin{align}\label{4.59}
        \widetilde{\Omega}_\mu&=\{x\in\Omega_\mu \mid
        G(\mu_j(x),x)=-\frac{y(\hat{\mu}_j(x))}{\mu_j(x)}\neq 0,\,j=0,\ldots,n\}
        \nonumber \\
        &=\{x\in\Omega_\mu \mid \mu_j(x)
        \notin \{E_m \}_{m=0,\ldots,2n+1},\,j=0,\ldots,n\},
      \end{align}
 one infers that
      \begin{equation}\label{4.60}
        P_{n}(\mu_j(x),x)=0,\quad
        j=0,\ldots,n,\, ~ x\in\widetilde{\Omega}_\mu.
      \end{equation}
 Since $P_{n}(z)$ has degree at most $n$, (\ref{4.60})
 implies
     \begin{equation}\label{4.61}
        P_{n}=0
        \quad \textrm{on $\mathbb{C}\times\widetilde{\Omega}_\mu$},
     \end{equation}
 and hence (\ref{2.15c}) holds, that is,
       \begin{equation}\label{4.62}
        z^2G_{n,x}(z)=-H_n(z)-(\rho^2+u_{xx}z)F_{n}(z)
         \quad
       \end{equation}
       $\textrm{on $\mathbb{C}\times\widetilde{\Omega}_\mu$}.$
Inserting (\ref{4.62}) and (\ref{4.48}) into (\ref{4.56}) yields
         \begin{equation}\label{4.63}
             F_{n}(z) (H_{n,x}(z)-2(\rho^2+u_{xx}z)G_n(z))=0,
         \end{equation}
 namely
       \begin{equation}\label{4.64}
        H_{n,x}(z)=2(\rho^2+u_{xx}z)G_n(z)
               \end{equation}
 \textrm{on $\mathbb{C}\times\widetilde{\Omega}_\mu$}.
 Thus, we obtain the fundamental equations (\ref{2.15a})-(\ref{2.15c})
 and (\ref{2.19})
 on $\mathbb{C}\times\widetilde{\Omega}_\mu$.
In order to extend these results to $\Omega_\mu$, we next
 investigate the case where $\hat{\mu}_j$ hits a
 branch point $(E_{m_0},0)$. Hence, we suppose
   \begin{equation}\label{4.65}
      \mu_{j_1}(x)\rightarrow E_{m_0}
      \quad \textrm{as $x\rightarrow x_0\in\Omega_\mu$}
   \end{equation}
 for some $j_1\in\{0,\ldots,n\},\,m_0\in\{0,\ldots,2n+1\}$.
 Introducing
\begin{equation}\label{4.66}
   \begin{split}
 & \zeta_{j_1}(x)=\sigma(\mu_{j_1}(x)-E_{m_0})^{1/2},
   \quad \sigma=\pm1,\quad \\
  & \mu_{j_1}(x)=E_{m_0}+\zeta_{j_1}(x)^2,
    \end {split}
 \end{equation}
 for some $x$ in an open interval centered near $x_0$,
 the Dubrovin
 equation (\ref{3.38}) for $\mu_{j_1}$ becomes
\begin{align}\label{4.67}
 \zeta_{j_1,x}(x)&=c(\sigma)
 \frac{ 2a }{E_{m_0}}
 \Big(\prod_{\scriptstyle m=0 \atop \scriptstyle m\neq
  m_0}^{2n+1}(E_{m_0}-E_m)\Big)^{1/2} \nonumber \\
   &~~~~~\times \prod_{\scriptstyle k=0 \atop \scriptstyle k\neq
   j_1}^{n}(E_{m_0}-\mu_k(x))^{-1}(1+O(
   \zeta_{j_1}(x)^2))
 \end{align}
 for some $| c(\sigma)|=1$. Hence, (\ref{4.61})-(\ref{4.64}) extend to
  $\Omega_\mu$ by continuity. We have now established relations
  (\ref{2.15a})-(\ref{2.15c}) on $\mathbb{C}\times \Omega_\mu$, and one can proceed
  as in Section 2 to obtain
  (\ref{4.47}). \quad $\square$

\newtheorem{rem4.11}[lem4.1]{Remark}
  \begin{rem4.11}
  Although we formulated Theorem $\ref{theorem4.10}$ in terms of
  $\{\mu_j\}_{j=0,\ldots,n}$ only, the analogous result (and strategy of proof)
  obviously works in terms of   $\{\nu_j\}_{j=0,\ldots,n}$.
  \end{rem4.11}

\newtheorem{rem4.12}[lem4.1]{Remark}
   \begin{rem4.12}\label{remark4.12}
    A closer look at Theorem $\ref{theorem4.10}$ reveals that
    $u,\rho$ are uniquely determined in an open neighborhood $\Omega$ of $x_0$
    by $\mathcal{K}_n$ and the initial condition
    $(\hat{\mu}_0(x_0),\hat{\mu}_1(x_0),\ldots,\hat{\mu}_n(x_0))$,
    or equivalently, by the  auxiliary divisor
     $\mathcal{D}_{\hat{\mu}_0(x_0) \underline{\hat{\mu}}(x_0)} $ at $x=x_0$.
         Conversely, given $\mathcal{K}_n$
    and $u,\rho$ in an open neighborhood $\Omega$ of $x_0$, one can
    construct the corresponding polynomial $F_{n}(z,x)$, $G_n(z,x)$
    and $H_n(z,x)$ for $x\in\Omega$, and then recover the auxiliary
    divisor $\mathcal{D}_{\hat{\mu}_0(x)\underline{\hat{\mu}}(x)}$ for
    $x\in\Omega$ from the zeros of $F_{n}(z,x)$ and from $(\ref{3.10})$.
    In this sense, once the curve $\mathcal{K}_n$ is fixed, elements of the
    isospectral class of the $HS2$ potentials $u,\rho$ can be characterized by
    nonspecial auxiliary divisors
    $\mathcal{D}_{\hat{\mu}_0(x) \underline{\hat{\mu}}(x)}$.
   \end{rem4.12}

\section{The time-dependent HS2 formalism}
 In this section, we
 extend the algebro-geometric analysis of Section 3 to the
 time-dependent HS2 hierarchy.

 Throughout this section, we assume (\ref{2.2}) holds.

 The time-dependent algebro-geometric initial value problem of the
 HS2 hierarchy is to solve the time-dependent $r$th HS2 flow with
 a stationary solution of the $n$th equation as initial data in the
 hierarchy. More precisely, given $n\in\mathbb{N}_0$, based on the
 solution $u^{(0)}, \rho^{(0)}$ of the $n$th stationary HS2 equation
 $\textrm{s-HS2}_n(u^{(0)},\rho^{(0)})=0$ associated with $\mathcal{K}_n$ and a
 set of integration constants $\{c_l\}_{l=1,\ldots,n} \subset
 \mathbb{C}$, we want to construct  a solution $u, \rho$ of the $r$th HS2 flow
 $\mathrm{HS2}_r(u,\rho)=0$ such that $u(t_{0,r})=u^{(0)},$  $\rho(t_{0,r})=\rho^{(0)},$ for some
 $t_{0,r} \in \mathbb{R},~r\in\mathbb{N}_0$.

 To emphasize that the integration
 constants in the definitions of the stationary and the time-dependent HS2
 equations are independent of each other, we indicate this by adding a tilde
 on all the time-dependent quantities. Hence, we employ the notation
 $\widetilde{V}_r,$ $\widetilde{F}_{r},$ $\widetilde{G}_r,$
 $\widetilde{H}_r,$ $\tilde{f}_{s}$, $\tilde{g}_{s},$ $\tilde{h}_{s}$, $\tilde{c}_s$
 in order to distinguish them from $V_n,$ $F_{n},$ $G_n,$ $H_n,$ $f_{l},$ $g_{l},$ $h_{l}$,
 $c_l$  in the following.
 In addition, we mark
 the individual $r$th HS2 flow by a separate time variable $t_r \in
 \mathbb{R}$.

 Summing up, we are seeking a solution $u, \rho$ of the time-dependent algebro-geometric initial
 value problem
   \begin{align}
    &
       \mathrm{HS2}_r(u,\rho)=
        \left(
          \begin{array}{c}
            -u_{xxt_r}+2u_{xx}\tilde{f}_{r+1,x}+u_{xxx}\tilde{f}_{r+1}
          +2\rho\rho_x \tilde{f}_{r}+2\rho^2\tilde{f}_{r,x} \\
            2\rho\rho_{t_r}-2\rho \rho_x \tilde{f}_{r+1}-2\rho^2 \tilde{f}_{r+1,x} \\
          \end{array}
        \right)=0, \label{5.1}
         \\
       & (u,\rho)|_{t_r=t_{0,r}}=(u^{(0)},\rho^{(0)}), \nonumber \\
     &
      \textrm{s-HS2}_n(u^{(0)},\rho^{(0)})=
     \left(
       \begin{array}{c}
         2u_{xx}f_{n+1,x}+u_{xxx}f_{n+1}
          +2\rho \rho_x f_{n}+2\rho^2 f_{n,x}  \\
         -2\rho \rho_x f_{n+1}-2\rho^2 f_{n+1,x} \\
       \end{array}
     \right)=0, \label{5.1a8}
    \end{align}
for some $t_{0,r}\in\mathbb{R},$ $n,r\in\mathbb{N}_0$, where $u=u(x,t_r),$ $\rho=\rho(x,t_r)$
 satisfy (\ref{2.2}), and the curve $\mathcal{K}_n$ is
 associated with the initial data $(u^{(0)}, \rho^{(0)})$ in (\ref{5.1a8}). Noticing that
 the HS2 flows are isospectral, we further
 assume that (\ref{5.1a8}) holds not only for $t_r=t_{0,r}$, but also for all $t_r \in
 \mathbb{R}$. Hence, we start with
 the zero-curvature equations
    \begin{equation}\label{5.3}
        U_{t_r}-\widetilde{V}_{r,x}+[U,\widetilde{V}_r]=0,
    \end{equation}
    \begin{equation}\label{5.4}
        -V_{n,x}+[U,V_n]=0,
    \end{equation}
 where
 \begin{equation}\label{5.5}
    \begin{split}
    & U(z)=
     \left(
       \begin{array}{cc}
         0 & 1 \\
         -z^{-2}\rho^2-z^{-1}u_{xx} & 0 \\
       \end{array}
     \right),
     \\
    & V_n(z)=
    \left(
      \begin{array}{cc}
      -G_n(z) & F_{n}(z) \\
      z^{-2}H_n(z) & G_n(z) \\
      \end{array}
    \right),
      \\
    & \widetilde{V}_r(z)=
       \left(
         \begin{array}{cc}
           -\widetilde{G}_r(z) & \widetilde{F}_{r}(z) \\
           z^{-2}\widetilde{H}_r(z) & \widetilde{G}_r(z) \\
         \end{array}
       \right),
    \end{split}
  \end{equation}
and
  \begin{eqnarray}\label{5.6}
   &&
    F_{n}(z)=\sum_{l=0}^{n+1} f_{l} z^{n+1-l}
    =f_0\prod_{j=0}^n (z-\mu_j),
    \\
   &&
    G_n(z)=\sum_{l=0}^{n} g_{l} z^{n-l},
     \\
   &&
    H_n(z)=\sum_{l=0}^{n+1} h_{l} z^{n+1-l}
    =h_0 \prod_{l=0}^{n} (z-\nu_l),\\
   &&
    \widetilde{F}_{r}(z)=\sum_{s=0}^{r+1} \tilde{f}_{s} z^{r+1-s},
    \\
   &&
    \widetilde{G}_r(z)=\sum_{s=0}^{r} \tilde{g}_{s} z^{r-s},
    \\
    &&
    \widetilde{H}_r(z)=\sum_{s=0}^{r+1} \tilde{h}_{s} z^{r+1-s},
   \end{eqnarray}
 for fixed $n,r\in \mathbb{N}_0$. Here, $\{f_{l}\}_{l=0,\ldots,n+1},$
 $\{g_{l}\}_{l=0,\ldots,n}$, $\{h_{l}\}_{l=0,\ldots,n+1}$,
  $\{\tilde{f}_{s}\}_{s=0,\ldots,r+1},$
 $\{\tilde{g}_{s}\}_{s=0,\ldots,r}$, and $\{\tilde{h}_{s}\}_{s=0,\ldots,r+1}$
 are defined as in
 (\ref{2.3}), with $u(x), \rho(x)$ replaced by $u(x,t_r), \rho(x,t_r)$ etc., and with appropriate
 integration constants.
 Explicitly, (\ref{5.3}) and (\ref{5.4}) are equivalent to
  \begin{align}
        &
      -2\rho \rho_{t_r}-zu_{xxt_r}-\widetilde{H}_{r,x}
        +2(\rho^2+zu_{xx})\widetilde{G}_r=0,  \label{5.12} \\
      &
       \widetilde{F}_{r,x} =2\widetilde{G}_r,\label{5.13} \\
      &
      z^2\widetilde{G}_{r,x}=-\widetilde{H}_r
        - (\rho^2+zu_{xx})\widetilde{F}_{r}\label{5.14}
    \end{align}
and
   \begin{align}
    &
     F_{n,x}=2G_n, \label{5.15} \\
    &
      H_{n,x}=2(\rho^2+zu_{xx})G_n, \label{5.16} \\
    &
    z^2G_{n,x}=-H_n -(\rho^2+zu_{xx})F_{n}.\label{5.17}
   \end{align}
From (\ref{5.15})-(\ref{5.17}), one finds
  \begin{equation}\label{5.18}
    \frac{d}{dx} \mathrm{det}(V_n(z))=-\frac{1}{z^2}\frac{d}{dx}
    \Big( z^2G_n(z)^2+F_{n}(z)H_n(z) \Big)=0,
  \end{equation}
and meanwhile (see Lemma \ref{lemma5.2})
  \begin{equation}\label{5.19}
    \frac{d}{dt_r} \mathrm{det}(V_n(z))=-\frac{1}{z^2}\frac{d}{dt_r}
    \Big( z^2G_n(z)^2+F_{n}(z)H_n(z) \Big)=0.
  \end{equation}
Hence, $z^2G_n(z)^2+F_{n}(z)H_n(z)$ is independent of variables
both $x$ and $t_r$, which implies the fundamental identity (\ref{2.19}) holds,
   \begin{equation}\label{5.20}
   z^2G_n(z)^2+F_{n}(z)H_n(z)=R_{2n+2}(z),
   \end{equation}
and the hyperelliptic curve
$\mathcal{K}_n$ is still given by (\ref{2.27}).

Next, we define the time-dependent Baker-Akhiezer function
$\psi(P,x,x_0,t_r,t_{0,r})$ on
$\mathcal{K}_n \setminus \{ P_{\infty_\pm},P_0 \}$ by
  \begin{equation}\label{5.21}
         \begin{split}
          & \psi(P,x,x_0,t_r,t_{0,r})=\left(
                            \begin{array}{c}
                              \psi_1(P,x,x_0,t_r,t_{0,r}) \\
                              \psi_2(P,x,x_0,t_r,t_{0,r}) \\
                            \end{array}
                          \right), \\
          & \psi_x(P,x,x_0,t_r,t_{0,r})=U(u(x,t_r), \rho(x,t_r), z(P))\psi(P,x,x_0,t_r,t_{0,r}),\\
          &  \psi_{t_r}(P,x,x_0,t_r,t_{0,r})=\widetilde{V}_r(u(x,t_r),\rho(x,t_r),z(P))
               \psi(P,x,x_0,t_r,t_{0,r}), \\
           & zV_n(u(x,t_r),\rho(x,t_r),z(P))\psi(P,x,x_0,t_r,t_{0,r})
           =y(P)\psi(P,x,x_0,t_r,t_{0,r}),\\
          & \psi_1(P,x_0,x_0,t_{0,r},t_{0,r})=1; \\
          &
          P=(z,y)\in \mathcal{K}_{n}
           \setminus \{P_{\infty_\pm},P_0\},~(x,t_r)\in
           \mathbb{R}^2.
         \end{split}
   \end{equation}
Closely related to $\psi(P,x,x_0,t_r,t_{0,r})$ is the following
meromorphic function $\phi(P,x,t_r)$ on $\mathcal{K}_{n}$ defined by
        \begin{equation}\label{5.23}
         \phi(P,x,t_r)=z
         \frac{ \psi_{1,x}(P,x,x_0,t_r,t_{0,r})}
          {\psi_1(P,x,x_0,t_r,t_{0,r})},
                      \quad P \in
          \mathcal{K}_{n}\setminus \{P_{\infty_\pm},P_0\},
          ~ (x,t_r)\in \mathbb{R}^2
        \end{equation}
such that
\begin{align}\label{5.22}
   \psi_1(P,x,x_0,t_r,t_{0,r})&=\mathrm{exp}\Big(
     \int_{t_{0,r}}^{t_r} ds
     (z^{-1}\widetilde{F}_{r}(z,x_0,s)\phi(P,x_0,s)
     \nonumber \\
     &~~~
     -\widetilde{G}_r(z,x_0,s))
     + z^{-1} \int_{x_0}^x dx^\prime
     \phi(P,x^\prime,t_r)
      \Big), \nonumber \\
     & ~~~~~~~~~~~
     P=(z,y)\in \mathcal{K}_{n}
           \setminus \{P_{\infty_\pm},P_0\}.
   \end{align}
Then, using (\ref{5.21}) and (\ref{5.23}),  one infers that
   \begin{align}\label{5.24}
        \phi(P,x,t_r)&= \frac{y+zG_n(z,x,t_r)}{F_{n}(z,x,t_r)}
           \nonumber \\
        &=
         \frac{H_n(z,x,t_r)}{y-zG_n(z,x,t_r)},
    \end{align}
 and
      \begin{equation}\label{5.25}
        \psi_2(P,x,x_0,t_r,t_{0,r})=
        \psi_1(P,x,x_0,t_r,t_{0,r})\phi(P,x,t_r)/z.
      \end{equation}
In analogy to (\ref{3.10}) and (\ref{3.11}), we introduce
     \begin{align}
      & \hat{\mu}_j(x,t_r)=(\mu_j(x,t_r),
       -\mu_j(x,t_r)G_n(\mu_j(x,t_r),x,t_r))
       \in \mathcal{K}_n,
        \label{5.26} \\
      &~~~~~~~~~~~~~~~~~~~~~~~~~~~~~~~~~~~~~~~~~
        j=0,\ldots,n, ~(x,t_r)\in \mathbb{R}^2,  \nonumber \\
      &
      \hat{\nu}_l(x,t_r)=(\nu_l(x,t_r),
       \nu_l(x,t_r)G_n(\nu_l(x,t_r),x,t_r))
       \in \mathcal{K}_n,
        \label{5.27} \\
      &~~~~~~~~~~~~~~~~~~~~~~~~~~~~~~~~~~~~~~~~~
       l=0,\ldots,n, ~(x,t_r)\in \mathbb{R}^2. \nonumber
     \end{align}
The regularity properties of $F_{n}$, $H_n$, $\mu_j$, and $\nu_l$ are
analogous to those in Section 3 due to assumptions
(\ref{2.2}).
Similar to (\ref{3.14}), the divisor $(\phi(P,x,t_r))$ of
$\phi(P,x,t_r)$ reads
 \begin{equation}\label{5.28}
           (\phi(P,x,t_r))=
           \mathcal{D}_{\hat{\nu}_0(x,t_r)\underline{\hat{\nu}}(x,t_r)}(P)
           -\mathcal{D}_{\hat{\mu}_0(x,t_r) \underline{\hat{\mu}}(x,t_r)}(P)
  \end{equation}
with
  \begin{equation}\label{5.29}
    \underline{\hat{\mu}}=\{\hat{\mu}_1,\ldots,\hat{\mu}_{n}\},
    \quad
    \underline{\hat{\nu}}=\{\hat{\nu}_1,\ldots,\hat{\nu}_{n}\}
    \in \mathrm{Sym}^n (\mathcal{K}_n).
  \end{equation}

The properties of $\phi(P,x,t_r)$ are summarized as follows.
\newtheorem{lem5.1}{Lemma}[section]
 \begin{lem5.1}
  Assume $(\ref{2.2})$ and suppose that $(\ref{5.3})$,
    $(\ref{5.4})$ hold. Moreover, let
    $P=(z,y) \in \mathcal{K}_{n}\setminus \{P_{\infty_\pm},P_0\}$ and
    $(x,t_r)\in \mathbb{R}^2.$ Then $\phi$ satisfies
    \begin{align}
    &
    \phi_x(P)+ z^{-1}\phi(P)^2 =-z^{-1}\rho^2-u_{xx},\label{5.30} \\
    &
    \phi_{t_r}(P)= (-z\widetilde{G}_r(z)+\widetilde{F}_{r}(z)\phi(P))_x \label{5.31} \\
    &~~~~~~~~
             = z^{-1}\widetilde{H}_r(z)+(z^{-1}\rho^2+ u_{xx})\widetilde{F}_{r}(z)
                +(\widetilde{F}_{r}(z)\phi(P))_x, \nonumber \\
    &
    \phi_{t_r}(P)=z^{-1}\widetilde{H}_r(z)+2\widetilde{G}_r(z)\phi(P)
         -z^{-1}\widetilde{F}_{r}(z)\phi(P)^2,\label{5.32}\\
         &\phi(P)\phi(P^\ast)=-\frac{H_n(z)}{F_{n}(z)},\label{5.33}\\
         &\phi(P)+\phi(P^\ast)=2\frac{zG_n(z)}{F_{n}(z)},\label{5.34}\\
         & \phi(P)-\phi(P^\ast)=\frac{2y}{F_{n}(z)}.\label{5.35}
   \end{align}
\end{lem5.1}
\noindent
{\it Proof.}~Equations (\ref{5.30}) and (\ref{5.33})-(\ref{5.35})
can be proved as in Lemma \ref{lemma3.1}. Using (\ref{5.21}) and
(\ref{5.23}), one infers that
\begin{align}\label{5.36}
       & \phi_{t_r}=
        z(\mathrm{ln}\psi_1)_{xt_r}=z(\mathrm{ln}\psi_1)_{t_rx}
        =z\Big(\frac{\psi_{1,t_r}}{\psi_1}\Big)_x
          \nonumber \\
      &~~~~
  = z
 \Big(\frac{-\widetilde{G}_r\psi_1+\widetilde{F}_{r}\psi_2}{\psi_1}\Big)_x
     = (-z\widetilde{G}_r+\widetilde{F}_{r}\phi)_x.
 \end{align}
 Insertion of (\ref{5.14}) into
(\ref{5.36}) then yields (\ref{5.31}).
To prove (\ref{5.32}),
 one observes that
 \begin{align}\label{5.37}
     \phi_{t_r} =& z \Big(\frac{\psi_2}{\psi_1}\Big)_{t_r}
     =
      z \Big(\frac{\psi_{2,t_r}}{\psi_1}-
      \frac{\psi_2\psi_{1,t_r}}{\psi_1^2} \Big)
        \nonumber \\
     =&
     z \Big(\frac{z^{-2}\widetilde{H}_r\psi_1+\widetilde{G}_r\psi_2}{\psi_1}
     -z^{-1}\phi\frac{-\widetilde{G}_r\psi_1+\widetilde{F}_{r}\psi_2}{\psi_1}\Big)
       \nonumber \\
     =&
     z^{-1}\widetilde{H}_r+2\widetilde{G}_r\phi-z^{-1}\widetilde{F}_{r}\phi^2,
 \end{align}
which leads to (\ref{5.32}). Alternatively, one can also insert (\ref{5.12})-(\ref{5.14})
into (\ref{5.31}) to obtain (\ref{5.32}). \quad $\square$ \\

Next, we determine the time evolution of $F_{n}$, $G_n$, and $H_n$,
using relations (\ref{5.12})-(\ref{5.14}) and (\ref{5.15})-(\ref{5.17}).
\newtheorem{lem5.2}[lem5.1]{Lemma}
 \begin{lem5.2}\label{lemma5.2}
  Assume $(\ref{2.2})$ and suppose that $(\ref{5.3})$,
  $(\ref{5.4})$ hold. Then
   \begin{align}
    & F_{n,t_r}=2(G_n\widetilde{F}_{r}-\widetilde{G}_rF_{n}), \label{5.38} \\
    &  z^2G_{n,t_r}=\widetilde{H}_rF_{n}-H_n\widetilde{F}_{r}, \label{5.39} \\
    &  H_{n,t_r}=2(H_n\widetilde{G}_r-G_n\widetilde{H}_r). \label{5.40}
   \end{align}
 Equations $(\ref{5.38})$--$(\ref{5.40})$ are equivalent to
    \begin{equation}\label{5.41}
        -V_{n,t_r}+[\widetilde{V}_r,V_n]=0.
    \end{equation}
 \end{lem5.2}
\noindent
{\it Proof.}~Differentiating (\ref{5.35}) with
respect to $t_r$ naturally yields
  \begin{equation}\label{5.42}
    (\phi(P)-\phi(P^\ast))_{t_r}=-2yF_{n,t_r}F_{n}^{-2}.
  \end{equation}
On the other hand, using (\ref{5.32}), (\ref{5.34}), and (\ref{5.35}),
the left-hand side of (\ref{5.42}) can be expressed as
   \begin{align}\label{5.43}
    \phi(P)_{t_r}-\phi(P^\ast)_{t_r}=&
    2\widetilde{G}_r(\phi(P)-\phi(P^\ast))-z^{-1}\widetilde{F}_{r}
    (\phi(P)^2-\phi(P^\ast)^2)
     \nonumber \\
    =&
    4y(\widetilde{G}_rF_{n}-\widetilde{F}_{r}G_n)F_{n}^{-2}.
   \end{align}
Combining (\ref{5.42}) and (\ref{5.43}) then proves (\ref{5.38}).
Similarly, differentiating (\ref{5.34}) with respect
to $t_r$, one finds
  \begin{equation}\label{5.44}
    (\phi(P)+\phi(P^\ast))_{t_r}=2z(G_{n,t_r}F_{n}-G_nF_{n,t_r})F_{n}^{-2}.
  \end{equation}
Meanwhile, the
left-hand side of (\ref{5.44})  also equals
  \begin{align}\label{5.45}
  \phi(P)_{t_r}+\phi(P^\ast)_{t_r}=&
   2\widetilde{G}_r(\phi(P)+\phi(P^\ast))
   -z^{-1}\widetilde{F}_{r}(\phi(P)^2+\phi(P^\ast)^2)+2z^{-1}\widetilde{H}_r
    \nonumber \\
   =&
   -2zG_nF_{n}^{-2}F_{n,t_r}
   +2z^{-1}F_{n}^{-1}(\widetilde{H}_rF_{n}-\widetilde{F}_{r}H_n),
  \end{align}
  using (\ref{5.32}), (\ref{5.33}), and (\ref{5.34}).
Equation (\ref{5.39}) is  clear from
(\ref{5.44}) and (\ref{5.45}). Then, (\ref{5.40}) follows by
differentiating (\ref{2.19}), that is, $z^2G_n^2+F_{n}H_n=R_{2n+2}(z)$,
with respect to $t_r$, and using  (\ref{5.38}) and (\ref{5.39}).
Finally,
a direct calculation shows (\ref{5.41}) holds. \quad $\square$
\vspace{0.1cm}

Basic properties of $\psi(P,x,x_0,t_r,t_{0,r})$ are summarized as follows.

\newtheorem{lem5.3}[lem5.1]{Lemma}
 \begin{lem5.3}
  Assume $(\ref{2.2})$ and suppose that $(\ref{5.3})$,
  $(\ref{5.4})$ hold.  Moreover, let
  $P=(z,y) \in \mathcal{K}_{n}\setminus \{P_{\infty_\pm},P_0\}$ and
   $(x,x_0,t_r,t_{0,r})\in \mathbb{R}^4.$
  Then, the Baker-Akhiezer function $\psi$ satisfies
  \begin{align}
   & \psi_1(P,x,x_0,t_r,t_{0,r}) =
    \Big(\frac{F_{n}(z,x,t_r)}{F_{n}(z,x_0,t_{0,r})}\Big)^{1/2}
    \mathrm{exp} \Bigg(
    \frac{y}{z}\int_{t_{0,r}}^{t_r} ds
    \widetilde{F}_{r}(z,x_0,s)F_{n}(z,x_0,s)^{-1}
    \nonumber \\
    &~~~~~~~~~~~~~~~~~~~~~~~~~~
    +\frac{y}{z} \int_{x_0}^x dx^\prime
    F_{n}(z,x^\prime,t_r)^{-1}
    \Bigg),\label{5.46} \\
  &  \psi_1(P,x,x_0,t_r,t_{0,r}) \psi_1(P^\ast,x,x_0,t_r,t_{0,r})
    =\frac{F_{n}(z,x,t_r)}{F_{n}(z,x_0,t_{0,r})}, \label{5.47} \\
  & \psi_2(P,x,x_0,t_r,t_{0,r}) \psi_2(P^\ast,x,x_0,t_r,t_{0,r})
    =-\frac{H_n(z,x,t_r)}{z^2 F_{n}(z,x_0,t_{0,r})}, \label{5.48} \\
  & \psi_1(P,x,x_0,t_r,t_{0,r}) \psi_2(P^\ast,x,x_0,t_r,t_{0,r})
    + \psi_1(P^\ast,x,x_0,t_r,t_{0,r}) \psi_2(P,x,x_0,t_r,t_{0,r})
    =2\frac{G_n(z,x,t_r)}{F_{n}(z,x_0,t_{0,r})},\label{5.49} \\
  & \psi_1(P,x,x_0,t_r,t_{0,r}) \psi_2(P^\ast,x,x_0,t_r,t_{0,r})
    - \psi_1(P^\ast,x,x_0,t_r,t_{0,r}) \psi_2(P,x,x_0,t_r,t_{0,r})
    =-\frac{2y}{z F_{n}(z,x_0,t_{0,r})}.\label{5.50}
  \end{align}
\end{lem5.3}
\noindent
{ \it Proof.}~To prove (\ref{5.46}), we first
consider the part of time variable in the definition (\ref{5.22}),
that is,
   \begin{equation}\label{5.51}
    \mathrm{exp} \left( \int_{t_{0,r}}^{t_r} ds~
    (z^{-1}\widetilde{F}_{r}(z,x_0,s)\phi(P,x_0,s)
    -\widetilde{G}_r(z,x_0,s))
    \right).
   \end{equation}
The integrand in the above integral equals
   \begin{align}\label{5.52}
   &
     z^{-1}\widetilde{F}_{r}(z,x_0,s)\phi(P,x_0,s)
     -\widetilde{G}_r(z,x_0,s)
     \nonumber \\
   &~~~~~
     =z^{-1}\widetilde{F}_{r}(z,x_0,s)
     \frac{y+zG_n(z,x_0,s)}{F_{n}(z,x_0,s)}
     -\widetilde{G}_r(z,x_0,s)
       \nonumber \\
     &~~~~~
     =\frac{y}{z}\widetilde{F}_{r}(z,x_0,s)F_{n}(z,x_0,s)^{-1}
      +(\widetilde{F}_{r}(z,x_0,s)G_n(z,x_0,s)
       \nonumber \\
    &~~~~~~~~
     -\widetilde{G}_r(z,x_0,s)F_{n}(z,x_0,s))F_{n}(z,x_0,s)^{-1}
       \nonumber \\
      &~~~~~
      =\frac{y}{z}\widetilde{F}_{r}(z,x_0,s)F_{n}(z,x_0,s)^{-1}
      +\frac{1}{2}\frac{F_{n,s}(z,x_0,s)}{F_{n}(z,x_0,s)},
   \end{align}
using (\ref{5.24}) and (\ref{5.38}). Hence,
(\ref{5.51}) can be expressed as
   \begin{equation}\label{5.53}
   \Big(\frac{F_{n}(z,x_0,t_r)}{F_{n}(z,x_0,t_{0,r})}\Big)^{1/2}
     \mathrm{exp} \left( \frac{y}{z} \int_{t_{0,r}}^{t_r} ds
     \widetilde{F}_{r}(z,x_0,s)F_{n}(z,x_0,s)^{-1}
     \right).
   \end{equation}
On the other hand, the
part of space variable in (\ref{5.22}) can be written as
 \begin{equation}\label{5.54}
    \Big(\frac{F_{n}(z,x,t_r)}{F_{n}(z,x_0,t_{r})}\Big)^{1/2}
    \mathrm{exp}\left(\frac{y}{z} \int_{x_0}^x dx^\prime
    F_{n}(z,x^\prime,t_r)^{-1}  \right),
 \end{equation}
 using the similar procedure in Lemma \ref{lemma3.2}. Then
combining (\ref{5.53}) and (\ref{5.54}) readily leads to (\ref{5.46}).
Evaluating (\ref{5.46}) at the points $P$ and $P^\ast$
and multiplying the resulting expressions yields (\ref{5.47}).
The remaining statements are direct consequences of (\ref{5.25}),
(\ref{5.33})-(\ref{5.35}), and (\ref{5.47}). \quad $\square$

\vspace{0.1cm}

In analogy to Lemma \ref{lemma3.4}, the dynamics of the zeros
$\{\mu_j(x,t_r)\}_{j=0,\ldots,n}$ and
$\{\nu_l(x,t_r)\}_{l=0,\ldots,n}$ of $F_{n}(z,x,t_r)$ and
$H_n(z,x,t_r)$ with respect to $x$ and $t_r$ are described in terms
of the following Dubrovin-type equations.

\newtheorem{lem5.4}[lem5.1]{Lemma}
 \begin{lem5.4}
   Assume $(\ref{2.2})$ and suppose that $(\ref{5.3})$,
   $(\ref{5.4})$ hold subject to the constraint $(\ref{2.26a})$.
\begin{itemize}
  \item[\emph{(i)}]
 Suppose that the zeros $\{\mu_j(x,t_r)\}_{j=0,\ldots,n}$
 of $F_{n}(z,x,t_r)$ remain distinct for $(x,t_r) \in
 \Omega_\mu,$ where $\Omega_\mu \subseteq \mathbb{R}^2$ is
  open
 and connected, then $\{\mu_j(x,t_r)\}_{j=0,\ldots,n}$ satisfy
 the system of differential equations,
    \begin{align}
        & \mu_{j,x}=4\frac{y(\hat{\mu}_j)}{\mu_j}
        \prod_{ \scriptstyle k=0 \atop \scriptstyle k \neq j}^n
        (\mu_j-\mu_k)^{-1},
        \quad j=0,\ldots,n, \label{5.56} \\
       &
       \mu_{j,t_r}=4\frac{\widetilde{F}_{r}(\mu_j) y(\hat{\mu}_j) }
                    {\mu_j}
         \prod_{ \scriptstyle k=0 \atop \scriptstyle k \neq j}^n
        (\mu_j-\mu_k)^{-1},
        \quad j=0,\ldots,n,\label{5.57}
    \end{align}
with initial conditions
       \begin{equation}\label{5.58}
         \{\hat{\mu}_j(x_0,t_{0,r})\}_{j=0,\ldots,n}
         \in \mathcal{K}_{n},
       \end{equation}
for some fixed $(x_0,t_{0,r}) \in \Omega_\mu$. The initial value
problem $(\ref{5.57})$, $(\ref{5.58})$ has a unique solution
satisfying
        \begin{equation}\label{5.59}
         \hat{\mu}_j \in C^\infty(\Omega_\mu,\mathcal{K}_{n}),
         \quad j=0,\ldots,n.
        \end{equation}

   \item[\emph{(ii)}]
 Suppose that the zeros $\{\nu_l(x,t_r)\}_{l=0,\ldots,n}$
 of $H_n(z,x,t_r)$ remain distinct for $(x,t_r) \in
 \Omega_\nu,$ where $\Omega_\nu \subseteq \mathbb{R}^2$ is open
 and connected, then
 $\{\nu_l(x,t_r)\}_{l=0,\ldots,n}$ satisfy the system of
 differential equations,
     \begin{align}
        & \nu_{l,x}=-2\frac{(\rho^2+u_{xx}\nu_l)y(\hat{\nu}_l)}{h_0~\nu_l}
        \prod_{ \scriptstyle k=0 \atop \scriptstyle k \neq l}^{n}
        (\nu_l-\nu_k)^{-1},
        \quad l=0,\ldots,n, \label{5.60} \\
        &
         \nu_{l,t_r}=2\frac{\widetilde{H}_r(\nu_l) y(\hat{\nu}_l) }
                     {h_0~\nu_l}
        \prod_{ \scriptstyle k=0 \atop \scriptstyle k \neq l}^{n}
        (\nu_l-\nu_k)^{-1},
        \quad l=0,\ldots,n, \label{5.61}
     \end{align}
with initial conditions
       \begin{equation}\label{5.62}
         \{\hat{\nu}_l(x_0,t_{0,r})\}_{l=0,\ldots,n}
         \in \mathcal{K}_{n},
       \end{equation}
for some fixed $(x_0,t_{0,r}) \in \Omega_\nu$. The initial value
problem $(\ref{5.61})$, $(\ref{5.62})$ has a unique solution
satisfying
        \begin{equation}\label{5.63}
         \hat{\nu}_l \in C^\infty(\Omega_\nu,\mathcal{K}_{n}),
         \quad l=0,\ldots,n.
        \end{equation}
  \end{itemize}
  \end{lem5.4}
\noindent
{\it Proof.}~It suffices to prove (\ref{5.57})
 since the argument for (\ref{5.61}) is analogous
 and that for (\ref{5.56}) and (\ref{5.60})
 has been given in the proof of Lemma \ref{lemma3.4}.
 Differentiating (\ref{5.6}) with respect to $t_r$ yields
     \begin{equation}\label{5.64}
        F_{n,t_r}(\mu_j)=-\frac{1}{2}\mu_{j,t_r}
         \prod_{ \scriptstyle k=0 \atop \scriptstyle k \neq j}^n
        (\mu_j-\mu_k).
     \end{equation}
On the other hand, inserting $z=\mu_j$ into (\ref{5.38}) and
using (\ref{5.26}), one finds
     \begin{equation}\label{5.65}
        F_{n,t_r}(\mu_j)=2G_n(\mu_j)\widetilde{F}_{r}(\mu_j)
         =2\frac{y(\hat{\mu}_j)}{-\mu_j}\widetilde{F}_{r}(\mu_j).
     \end{equation}
Combining (\ref{5.64}) and (\ref{5.65}) then yields (\ref{5.57}).
The rest is analogous to the proof of Lemma \ref{lemma3.4}.
\quad $\square$
\vspace{0.1cm}

Since the stationary trace formulas for HS2 invariants in terms of symmetric
functions of $\mu_j$ in Lemma \ref{lemma3.5} extend line by line to the corresponding
time-dependent setting, we next record the $t_r$-dependent trace formulas without
proof. For simplicity, we confine ourselves to the simplest one only.

\newtheorem{lem5.5}[lem5.1]{Lemma}
 \begin{lem5.5}
  Assume $(\ref{2.2})$, suppose that $(\ref{5.3})$, $(\ref{5.4})$ hold, and
  let $(x,t_r) \in \mathbb{R}^2$.
  Then,
    \begin{equation}\label{5.66}
     u(x,t_r)=\frac{1}{2}\sum_{j=0}^n \mu_j(x,t_r)-\frac{1}{2}\sum_{m=0}^{2n+1}
     E_m.
    \end{equation}
 \end{lem5.5}

\section{Time-dependent algebro-geometric solutions of HS2 hierarchy}
  In our final section, we extend the results of section 4 from the stationary
  HS2 hierarchy,  to the time-dependent case.
  We obtain Riemann theta function representations for
  the meromorphic function $\phi$, and especially, for the
  algebro-geometric solutions $u, \rho$ of the whole HS2 hierarchy.

 We first record the asymptotic properties of $\phi$ in the
  time-dependent case.

 \newtheorem{lem6.1}{Lemma}[section]
   \begin{lem6.1}\label{lemma6.1}
     Assume $(\ref{2.2})$ and suppose that $(\ref{5.3})$,
     $(\ref{5.4})$ hold. Moreover, let $P=(z,y)
    \in \mathcal{K}_n \setminus \{P_{\infty_\pm},P_0\}$, $(x,t_r) \in \mathbb{R}^2$. Then,
     \begin{align}
      &
       \phi(P)\underset{\zeta \rightarrow 0}{=}
       -u_x(x,t_r)+O(\zeta), \quad P \rightarrow P_{\infty_\pm}, \quad \zeta=z^{-1},
        \label{6.1} \\
      &
       \phi(P)\underset{\zeta \rightarrow 0}{=} i\rho(x,t_r)
       +\frac{iu_{xx}(x,t_r)-\rho_x(x,t_r)}{2\rho(x,t_r)}\zeta+O(\zeta^2),
       \quad P \rightarrow P_0, \quad \zeta=z.\label{6.2}
     \end{align}
   \end{lem6.1}
Since the proof of Lemma \ref{lemma6.1} is identical to the corresponding stationary results
in Lemma \ref{lemma4.1}, we omit the corresponding details.

\vspace{0.1cm}

Next, we investigate the properties of the Abel map. To do this,
let $\bar{\mu}=(\mu_0,\ldots,\mu_n) \in \mathbb{C}^{n+1}$, we define
the following symmetric functions by
\begin{equation}\label{6.4}
        \Psi_{k+1}(\bar{\mu})=(-1)^{k+1} \sum_{\underline{l}\in \mathcal{S}_{k+1}}
        \mu_{l_1}\ldots \mu_{l_{k+1}},
        \quad
        k=0,\ldots,n,
    \end{equation}
where $\mathcal{S}_{k+1}=\{\underline{l}=(l_1,\ldots,l_{k+1}) \in \mathbb{N}_0^{k+1}
        ~|~l_1 < \cdots < l_{k+1} \leq n\}$;
\begin{equation}\label{6.5}
 \Phi_{k+1}^{(j)}(\bar{\mu})=(-1)^{k+1} \sum_{\underline{l}\in \mathcal{T}_{k+1}^{(j)}}
           \mu_{l_1}\ldots \mu_{l_{k+1}}, \quad
 k=0,\ldots,n-1,
  \end{equation}
where  $\mathcal{T}_{k+1}^{(j)}=\{\underline{l} =(l_1,\ldots,l_{k+1}) \in
       \mathcal{S}_{k+1}
      ~|~ l_m \neq j\},$ ~ $ j=0,\ldots,n.$
For the
properties of $\Psi_{k+1}(\bar{\mu})$ and
$\Phi_{k+1}^{(j)}(\bar{\mu})$, we refer to Appendix E \cite{15}.

Introducing
 \begin{equation}\label{6.7}
     \tilde{d}_{r+1,k}(\underline{E})
     =\sum_{s=0}^{r+1-k} \tilde{c}_{r+1-k-s} \hat{c}_s(\underline{E}),
     \quad k=0,\ldots, r+1\wedge n+1,
   \end{equation}
for a given set of constants
$ \{ \tilde{c}_l\} _{l=1,\ldots,r+1} \subset \mathbb{C}$,
the corresponding homogeneous and nonhomogeneous quantities
$\widehat{F}_{r}(\mu_j)$ and $ \widetilde{F}_{r}(\mu_j)$
in the HS2 case are then given by
 \footnote{$m \wedge n = \mathrm{min}\{m,n\}$,
 $m \vee n =\mathrm{max} \{m,n\}$}
\begin{equation}\label{6.6}
      \begin{split}
     & \widehat{F}_{r}(\mu_j)= \sum_{s=(r-n)\vee 0}^{r+1}
      \hat{c}_s(\underline{E})
      \Phi_{r+1-s}^{(j)}(\bar{\mu}),
       \\
  & \widetilde{F}_{r}(\mu_j)= \sum_{s=0}^{r+1} \tilde{c}_{r+1-s}\widehat{F}_s(\mu_j)
  =
  \sum_{k=0}^{(r+1) \wedge (n+1)} \tilde{d}_{r+1,k}(\underline{E})
  \Phi_k^{(j)}(\bar{\mu}), \quad
    r\in \mathbb{N}_0, ~\tilde{c}_0=1,
     \end{split}
    \end{equation}
using (D.59) and (D.60) \cite{15}. Here, $\hat{c}_s(\underline{E})$, $s \in \mathbb{N}_0$,
is defined by (D.2) \cite{15}.
\vspace{0.1cm}

We now state the analog of Theorem \ref{theorem4.3}, which indicates marked differences between the HS2
hierarchy and other completely integrable systems such as the KdV and AKNS hierarchies.

\newtheorem{the6.2}[lem6.1]{Theorem}
 \begin{the6.2}
   Assume $(\ref{2.26a})$
   and suppose that $\{\hat{\mu}_j\}_{j=0,\ldots,n}$
   satisfies the Dubrovin equations $(\ref{5.56})$, $(\ref{5.57})$
   on an open set $\Omega_\mu \subseteq \mathbb{R}^2$ such that
   $\mu_j$, $j=0,\ldots,n,$ remain distinct and nonzero on $\Omega_\mu$
   and that $\widetilde{F}_{r}(\mu_j)
   \neq 0$ on $\Omega_\mu$, $ j=0,\ldots,n$. Introducing the associated divisor
   $\mathcal{D}_{\hat{\mu}_0(x,t_r)\underline{\hat{\mu}}(x,t_r)}$,
   one computes,
   \begin{align}
                   \partial_x
            \underline{\alpha}_{Q_0}(\mathcal{D}_{\hat{\mu}_0(x,t_r) \underline{\hat{\mu}}(x,t_r)})
            &=-\frac{4a}{ \Psi_{n+1}(\bar{\mu}(x,t_r))}
             \underline{c}(1),
            \quad    (x,t_r) \in \Omega_\mu,  \label{6.8}
        \\
         \partial_{t_r}
         \underline{\alpha}_{Q_0}(\mathcal{D}_{\hat{\mu}_0(x,t_r)  \underline{\hat{\mu}}(x,t_r)})
        & =-\frac{4a}{ \Psi_{n+1}(\bar{\mu}(x,t_r))}
         \left( \sum_{k=0}^{(r+1) \wedge (n+1)} \tilde{d}_{r+1,k}(\underline{E})
          \Psi_k(\bar{\mu}(x,t_r)) \right) \underline{c}(1)
          \nonumber \\
         &+
         4a \left( \sum_{\ell=1 \vee (n+1-r)}^{n+1}
         \tilde{d}_{r+1,n+2-\ell}(\underline{E})
         \underline{c}(\ell)
         \right),
          \quad
         (x,t_r) \in \Omega_\mu. \label{6.9}
        \end{align}
 In particular, the Abel map dose not linearize the divisor
 $\mathcal{D}_{\hat{\mu}_0(x,t_r)\underline{\hat{\mu}}(x,t_r)}$ on $\Omega_\mu$.
 \end{the6.2}
\noindent
{\it Proof.}~Let $ (x,t_r) \in \Omega_{\mu}$.
It suffices to prove (\ref{6.9}), since (\ref{6.8})
is proved as in the stationary context of Theorem
\ref{theorem4.3}. We first recall a fundamental identity (E.10) \cite{15},
that is,
 \begin{equation}\label{6.10}
     \Phi_{k+1}^{(j)}(\bar{\mu})=\mu_j \Phi_{k}^{(j)}(\bar{\mu})
     +\Psi_{k+1}(\bar{\mu}),
     \quad k=0,\ldots,n, ~
     j=0,\ldots,n.
    \end{equation}
Then, applying (\ref{4.16}), (\ref{6.6}), and (\ref{6.10}),  one finds
    \begin{align}\label{6.11}
    \frac{\widetilde{F}_{r}(\mu_j)}{\mu_j}
     &=\mu_j^{-1}
     \sum_{m=0}^{(r+1) \wedge (n+1)} \tilde{d}_{r+1,m}(\underline{E})
     \Phi_m^{(j)}(\bar{\mu})
       \\
     &=
     \mu_j^{-1}
     \sum_{m=0}^{(r+1) \wedge (n+1)} \tilde{d}_{r+1,m}(\underline{E})
     \Big(\mu_j\Phi_{m-1}^{(j)}(\bar{\mu})
     +\Psi_{m}(\bar{\mu})\Big)
       \nonumber \\
     &=
    \sum_{m=1}^{(r+1) \wedge (n+1)} \tilde{d}_{r+1,m}(\underline{E})
    \Phi_{m-1}^{(j)}(\bar{\mu})
    -\sum_{m=0}^{(r+1) \wedge (n+1)} \tilde{d}_{r+1,m}(\underline{E})
    \Psi_{m}(\bar{\mu})
    \frac{\Phi_{n}^{(j)}(\bar{\mu})}{\Psi_{n+1}(\bar{\mu})}.\nonumber
    \end{align}
Hence, using  (\ref{5.57}), (\ref{6.11}), (E.4), (E.13), and (E.14)
\cite{15}, one infers that
   \begin{align}
   &
    \partial_{t_r} \Big(\sum_{j=0}^n \int_{Q_0}^{\hat{\mu}_j}\underline{\omega}\Big)
    =\sum_{j=0}^n \mu_{j,t_r} \sum_{k=1}^n
    \underline{c}(k)\frac{a~\mu_j^{k-1}}{y(\hat{\mu}_j)}
      \nonumber \\
    &=
    4a\sum_{j=0}^n \sum_{k=1}^n \underline{c}(k)
    \frac{\mu_j^{k-1}}{\prod_{\scriptstyle l=0 \atop \scriptstyle l \neq j}^n (\mu_j-\mu_l)}
    \frac{\widetilde{F}_{r}(\mu_j)}{\mu_j}
      \nonumber \\
    &=
     4a\sum_{j=0}^n \sum_{k=1}^n \underline{c}(k)
    \frac{\mu_j^{k-1}}{\prod_{\scriptstyle l=0 \atop \scriptstyle l \neq j}^n (\mu_j-\mu_l)}
         \nonumber \\
    &\times
     \Big(
      -\sum_{m=0}^{(r+1) \wedge (n+1)} \tilde{d}_{r+1,m}(\underline{E})
    \Psi_{m}(\bar{\mu})
    \frac{\Phi_{n}^{(j)}(\bar{\mu})}{\Psi_{n+1}(\bar{\mu})}
    +\sum_{m=1}^{(r+1) \wedge (n+1)} \tilde{d}_{r+1,m}(\underline{E})
    \Phi_{m-1}^{(j)}(\bar{\mu}) \Big)
      \nonumber \\
       &=
     -4a\sum_{m=0}^{(r+1) \wedge (n+1)} \tilde{d}_{r+1,m}(\underline{E})
     \frac{\Psi_{m}(\bar{\mu})}{\Psi_{n+1}(\bar{\mu})}
     \sum_{k=1}^n \sum_{j=0}^n \underline{c}(k)
     (U_{n+1}(\bar{\mu}))_{k,j}(U_{n+1}(\bar{\mu}))_{j,1}^{-1}
       \nonumber \\
     &~~~~~
     +4a \sum_{m=1}^{(r+1) \wedge (n+1)} \tilde{d}_{r+1,m}(\underline{E})
     \sum_{k=1}^n \sum_{j=0}^n \underline{c}(k)
     (U_{n+1}(\bar{\mu}))_{k,j}(U_{n+1}(\bar{\mu}))_{j,n-m+2}^{-1}
       \nonumber \\
     &=
     -\frac{4a}{\Psi_{n+1}(\bar{\mu})}
     \sum_{m=0}^{(r+1) \wedge (n+1)} \tilde{d}_{r+1,m}(\underline{E})
     \Psi_{m}(\bar{\mu}) \underline{c}(1)
     +4a\sum_{m=1}^{(r+1) \wedge (n+1)} \tilde{d}_{r+1,m}(\underline{E})
     \underline{c}(n-m+2)
        \nonumber \\
     &=
     -\frac{4a}{\Psi_{n+1}(\bar{\mu})}
     \sum_{m=0}^{(r+1) \wedge (n+1)} \tilde{d}_{r+1,m}(\underline{E})
     \Psi_{m}(\bar{\mu}) \underline{c}(1)
     +4a\sum_{m=1\vee (n+1-r) }^{ n+1} \tilde{d}_{r+1,n+2-m}(\underline{E})
     \underline{c}(m), \label{6.12}
   \end{align}
which is equivalent to (\ref{6.9}). \quad $\square$
\vspace{0.1cm}

The analogous results hold for the corresponding divisor
$\mathcal{D}_{\hat{\nu}_0(x,t_r)\underline{\hat{\nu}}(x,t_r)}$ associated with
$\phi(P,x,t_r)$.
\vspace{0.1cm}

 Next, recalling  the definition of
 $\underline{\widehat{B}}_{Q_0}$ and $\underline{\hat{\beta}}_{Q_0}$
 in (\ref{4.20}) and (\ref{4.21}), one obtains the following result.

 \newtheorem{the6.3}[lem6.1]{Corollary}
  \begin{the6.3}
   Assume $(\ref{2.26a})$  and suppose that $\{\hat{\mu}_j\}_{j=0,\ldots,n}$
   satisfies the Dubrovin equations $(\ref{5.56})$, $(\ref{5.57})$
   on an open set $\Omega_\mu \subseteq \mathbb{R}^2$ such that
   $\mu_j$, $j=0,\ldots,n,$ remain distinct and nonzero on $\Omega_\mu$
   and that $\widetilde{F}_{r}(\mu_j)
   \neq 0$ on $\Omega_\mu,$ $ j=0,\ldots,n$.
   Then, one computes
   \begin{align}
       & \partial_x \sum_{j=0}^n \int_{Q_0}^{\hat{\mu}_j(x,t_r)} \eta_1
        = -\frac{4a}{\Psi_{n+1}(\bar{\mu}(x,t_r))},
        \quad (x,t_r) \in \Omega_\mu, \label{6.13} \\
      &
        \partial_x \underline{\hat{\beta}}
        (\mathcal{D}_{\underline{\hat{\mu}}(x,t_r)})
        =
        \begin{cases}
         4a,
         ~~~~~~~~n=1,\\
          4a (0,\ldots,0,1),
          ~~~~~~~~n\geq 2,
        \end{cases}
        \quad  (x,t_r) \in \Omega_\mu, \label{6.14} \\
   &
     \partial_{t_r} \sum_{j=0}^n \int_{Q_0}^{\hat{\mu}_j(x,t_r)} \eta_1
     =-\frac{4a}{\Psi_{n+1}(\bar{\mu}(x,t_r))}
     \sum_{k=0}^{(r+1) \wedge (n+1)} \tilde{d}_{r+1,k}(\underline{E})
     \Psi_{k}(\bar{\mu}(x,t_r))
      \nonumber \\
      &~~~~~~~~~~~~~~~~~~~~~~~~+
     4a \tilde{d}_{r+1,n+1}(\underline{E}) \delta_{n+1,r+1\wedge n+1},
     \quad  (x,t_r) \in \Omega_\mu, \label{6.15} \\
   &
   \partial_{t_r} \underline{\hat{\beta}}
        (\mathcal{D}_{\underline{\hat{\mu}}(x,t_r)})
        =4a\Big(
     \sum_{s=0}^{r+1} \tilde{c}_{r+1-s}\hat{c}_{s+1-n}(\underline{E}),
     \ldots,
     \sum_{s=0}^{r+1} \tilde{c}_{r+1-s}\hat{c}_{s+1}(\underline{E}),
           \sum_{s=0}^{r+1} \tilde{c}_{r+1-s}\hat{c}_{s}(\underline{E})
        \Big),
         \nonumber \\
     & ~~~~~~~~~~~~~~~~~~~~~~
     \quad \hat{c}_{-l}(\underline{E})=0,~ l\in
     \mathbb{N},
     \quad  (x,t_r) \in \Omega_\mu.\label{6.16}
   \end{align}
  \end{the6.3}
\noindent
{ \it Proof.}~Equations (\ref{6.13}) and (\ref{6.14}) are proved as
in the stationary context of Corollary \ref{corollary4.4}.
Equation (\ref{6.15}) is a special case of (\ref{6.9}),
and (\ref{6.16}) follows by (\ref{6.12}), taking into account
(E.4) \cite{15}. \quad $\square$.

\vspace{0.2cm}

The fact that the Abel map does not effect a linearization of the divisor
$\mathcal{D}_{\hat{\mu}_0(x,t_r)\underline{\hat{\mu}}(x,t_r)}$ in the
time-dependent HS2 context, which is well known and discussed
(using different approaches)
by Constantin and McKean \cite{C1}, Alber, Camassa, Fedorov, Holm,
and Marsden \cite{A1}, Alber and Fedorov \cite{A2, A3}.
The change of variables
    \begin{equation} \label{6.17}
             x \mapsto \tilde{x}= \int^x dx^\prime
        \left(\frac{4a}{\Psi_{n+1}(\bar{\mu}(x^\prime,t_r))}
        \right)
    \end{equation}
and
    \begin{align}
      t_r \mapsto  \tilde{t}_r &=\int^{t_r} ds
     \Bigg(
     \frac{4a}{ \Psi_{n+1}(\bar{\mu}(x,s))}
          \sum_{k=0}^{(r+1) \wedge (n+1)} \tilde{d}_{r+1,k}(\underline{E})
          \Psi_k(\bar{\mu}(x,s))
         \nonumber \\
         & ~~~~~
         -4a \sum_{\ell=1 \vee (n+1-r)}^{n+1}
         \tilde{d}_{r+1,n+2-\ell}(\underline{E})
         \frac{\underline{c}(\ell)}{\underline{c}(1)}
              \Bigg)\label{6.18}
    \end{align}
linearizes the Abel map
$\underline{A}_{Q_0}(\mathcal{D}_{\hat{\tilde{\mu}}_0(\tilde{x},\tilde{t}_r)
\underline{\hat{\tilde{\mu}}}(\tilde{x},\tilde{t}_r)})$,
$\tilde{\mu}_j(\tilde{x},\tilde{t}_r)=\mu_j(x,t_r)$, $j=0,\ldots,n$.
The intricate relation between the variables $(x,t_r)$ and
$(\tilde{x},\tilde{t}_r)$ is detailed in (\ref{6.21}).
Our approach follows a route similar to Gesztesy and Holden's
treatment of the CH hierarchy \cite{15}.
\vspace{0.1cm}

Next, we shall provide the explicit representations of $\phi$ and $u,\rho$
in terms of the Riemann theta function associated with
$\mathcal{K}_n$, assuming the affine part of $\mathcal{K}_n$ to be
nonsingular.
Recalling (\ref{4.26})-(\ref{4.32}), the analog of Theorem \ref{theorem4.5} in
the stationary case then reads as follows.

\newtheorem{the6.4}[lem6.1]{Theorem}
 \begin{the6.4}
   Assume $(\ref{2.2})$ and suppose that $(\ref{5.3})$,
   $(\ref{5.4})$ hold on $\Omega$ subject to the constraint
   $(\ref{2.26a})$. In addition,
   let $P=(z,y) \in \mathcal{K}_n \setminus \{P_0\}$
   and $(x,t_r),(x_0,t_{0,r}) \in \Omega$,  where $\Omega \subseteq
   \mathbb{R}^2$ is open and connected. Moreover, suppose that
   $\mathcal{D}_{\underline{\hat{\mu}}(x,t_r)}$, or equivalently,
   $\mathcal{D}_{\underline{\hat{\nu}}(x,t_r)}$ is nonspecial for
   $(x,t_r) \in \Omega$. Then, $\phi$, $u$, and $\rho$ admit the
   representations
   \begin{align}
    &
    \phi(P,x,t_r)= i \rho(x,t_r)
    \frac{\theta(\underline{z}(P,\underline{\hat{\nu}}(x,t_r) ))
         \theta(\underline{z}(P_0,\underline{\hat{\mu}}(x,t_r) ))}
     {\theta(\underline{z}(P_0,\underline{\hat{\nu}}(x,t_r) ))
     \theta(\underline{z}(P,\underline{\hat{\mu}}(x,t_r) ))}
     \mathrm{exp} \left(e_0-\int_{Q_0}^P \omega_{\hat{\mu}_0(x,t_r),\hat{\nu}_0(x,t_r) }^{(3)}
     \right), \label{6.19}
     \\
   &
   u(x,t_r)=-\frac{1}{2}\sum_{m=0}^{2n+1} E_m
             +\frac{1}{2}\sum_{j=1}^n \lambda_j
    -\frac{1}{2}\sum_{j=1}^n U_j \partial_{\omega_j}
              \mathrm{ln}
              \left(
    \frac{\theta(\underline{z}(P_{\infty_+},\underline{\hat{\mu}}(x,t_r) )+\underline{\omega})}
         {\theta(\underline{z}(P_{\infty_-},\underline{\hat{\mu}}(x,t_r) )+\underline{\omega})}
              \right)
              \Big|_{\underline{\omega}=0},\label{6.20}
    \\
    &
    \rho(x,t_r)=i u_x(x,t_r)
    \frac{\theta(\underline{z}(P_0,\underline{\hat{\nu}}(x,t_r) ))
         \theta(\underline{z}(P_{\infty_+},\underline{\hat{\mu}}(x,t_r) ))}
     {\theta(\underline{z}(P_{\infty_+},\underline{\hat{\nu}}(x,t_r) ))
     \theta(\underline{z}(P_0,\underline{\hat{\mu}}(x,t_r) ))}.\label{6.20a18}
   \end{align}
Moreover, let $\widetilde{\Omega} \subseteq \Omega$ be such that
$\mu_j$, $j=0,\ldots,n,$ are nonvanishing on $\widetilde{\Omega}$.
Then, the constraint
    \begin{align}
     &
     4a(x-x_0)
     +4a(t_r-t_{0,r}) \sum_{s=0}^{r+1} \tilde{c}_{r+1-s} \hat{c}_s(\underline{E})
     =\sum_{j=1}^n \left( \int_{a_j} \tilde{\omega}_{P_{\infty_+}P_{\infty_-}}^{(3)}
     \right) c_j(1)
      \nonumber \\
     & ~~~~~~ \times~
     \left(-4a \int_{x_0}^x \frac{dx^\prime}{\prod_{k=0}^n \mu_k(x^\prime, t_r)}
          -4a \sum_{k=0}^{(r+1) \wedge (n+1)} \tilde{d}_{r+1,k}(\underline{E})
     \int_{t_{0,r}}^{t_r} \frac{\Psi_k(\bar{\mu}(x_0,s))}{\Psi_{n+1}(\bar{\mu}(x_0,s))}
     ds
     \right) \nonumber
     \\
    &~~~~~~+~
     4a(t_r-t_{0,r}) \sum_{\ell=1 \vee (n+1-r) }^{n+1}
     \tilde{d}_{r+1,n+2-\ell}(\underline{E})
     \sum_{j=1}^n \left( \int_{a_j} \tilde{\omega}_{P_{\infty_+}P_{\infty_-}}^{(3)}
     \right) c_j(\ell)
      \nonumber \\
     & ~~~~~~+~
     \mathrm{ln} \left(
     \frac{\theta(\underline{z}(P_{\infty_+},\underline{\hat{\mu}}(x,t_r) ))
           \theta(\underline{z}(P_{\infty_-},\underline{\hat{\mu}}(x_0,t_{0,r}) ))}
          {\theta(\underline{z}(P_{\infty_-},\underline{\hat{\mu}}(x,t_{r})))
          \theta(\underline{z}(P_{\infty_+},\underline{\hat{\mu}}(x_0,t_{0,r}) ))}
          \right), \nonumber \\
    &~~~~~~~~~~~~~~~~~~~~~~~~~~~~~~~~~~~~~~~~~~~~~
        (x,t_r), (x_0,t_{0,r}) \in \widetilde{\Omega} \label{6.21}
           \end{align}
holds, with
  \begin{align}
   &\underline{\hat{\alpha}}_{Q_0}(\mathcal{D}_{  \hat{\mu}_0(x,t_r) \underline{\hat{\mu}}(x,t_r)})
      =\underline{\hat{\alpha}}_{Q_0}(\mathcal{D}_{ \hat{\mu}_0(x_0,t_r)  \underline{\hat{\mu}}(x_0,t_r)})
   -4a\left(   \int_{x_0}^x
   \frac{dx^\prime}{ \Psi_{n+1}(\bar{\mu}(x^\prime,t_r))} \right)
             \underline{c}(1) \label{6.22}
   \\
   &~~
   =\underline{\hat{\alpha}}_{Q_0}(\mathcal{D}_{ \hat{\mu}_0(x,t_{0,r})  \underline{\hat{\mu}}(x,t_{0,r})})
   -4a \left(
       \sum_{k=0}^{(r+1) \wedge (n+1)} \tilde{d}_{r+1,k}(\underline{E})
       \int_{t_{0,r}}^{t_r}
       \frac{\Psi_k(\bar{\mu}(x,s))}{\Psi_{n+1}(\bar{\mu}(x,s))}
       ds \right)\underline{c}(1)
   \nonumber \\
  &~~+
       4a(t_r-t_{0,r}) \Bigg(
       \sum_{\ell=1 \vee (n+1-r) }^{n+1}
       \tilde{d}_{r+1,n+2-\ell}(\underline{E})
       \underline{c}(\ell)
       \Bigg), \label{6.22a18}\\
 &~~~~~~~~~~~~~~~~~~~~~~~~~~~~~~~~~~~~~~~~~~~~~
  (x,t_r), (x_0,t_{0,r}) \in \widetilde{\Omega}. \nonumber
  \end{align}
\end{the6.4}
\noindent
{\it Proof.}~We first assume that $\mu_j$, $j=0,\ldots,n$,
are distinct and nonvanishing on $\widetilde{\widetilde{\Omega}}$ and
$\widetilde{F}_{r}(\mu_j) \neq 0$ on $\widetilde{\widetilde{\Omega}}$,
$j=0,\ldots,n,$ where $\widetilde{\widetilde{\Omega}} \subseteq \Omega$. Then,
the representation (\ref{6.19}) for $\phi$ on $\widetilde{\widetilde{\Omega}}$
follows by combining (\ref{5.28}), (\ref{6.1}), (\ref{6.2}), and
Theorem A.26 \cite{15}. The representation (\ref{6.20}) for $u$ on
$\widetilde{\widetilde{\Omega}}$ follows from the trace formulas (\ref{5.66})
and (F.59) \cite{15}. The representation (\ref{6.20a18}) for $\rho$ on
$\widetilde{\widetilde{\Omega}}$ is clear from (\ref{6.19}) and (\ref{6.1}).
In fact, since the proofs of (\ref{6.19}), (\ref{6.20}), and (\ref{6.20a18})
are identical to the corresponding stationary results
in Theorem \ref{theorem4.5}, which can be extended line by line to the
time-dependent setting, here we omit the corresponding details.
By continuity, (\ref{6.19}), (\ref{6.20}), and (\ref{6.20a18}) extend from
$\widetilde{\widetilde{\Omega}}$ to $\Omega$.
The
constraint (\ref{6.21}) then holds on $\widetilde{\widetilde{\Omega}}$ by
combining (\ref{6.13})-(\ref{6.16}) and (F.58) \cite{15}. Equations
(\ref{6.22}) and (\ref{6.22a18}) are clear from (\ref{6.8}) and (\ref{6.9}).
Again by continuity, (\ref{6.21})-(\ref{6.22a18}) extend from
 $\widetilde{\widetilde{\Omega}}$ to
$\widetilde{\Omega}$. \quad $\square$

\newtheorem{rem6.5}[lem6.1]{Remark}
 \begin{rem6.5}
  One observes  that
  $(\ref{6.22})$ and $(\ref{6.22a18})$ are equivalent to
  \begin{align}\label{6.24}
  \underline{\hat{\alpha}}_{Q_0}(\mathcal{D}_{  \hat{\mu}_0(x,t_r)\underline{\hat{\mu}}(x,t_r)})
               =&
  \underline{\hat{\alpha}}_{Q_0}(\mathcal{D}_{ \hat{\mu}_0(x_0,t_r) \underline{\hat{\mu}}(x_0,t_r)})
       -\underline{c}(1)(\tilde{x}-\tilde{x}_0)
          \\
        =&
  \underline{\hat{\alpha}}_{Q_0}(\mathcal{D}_{ \hat{\mu}_0(x,t_{0,r}) \underline{\hat{\mu}}(x,t_{0,r})})
        -\underline{c}(1)(\tilde{t}_r-\tilde{t}_{0,r}),
  \end{align}
  under the change of variables $x\mapsto \tilde{x}$ and $t_r \mapsto
  \tilde{t}_r$ in $(\ref{6.17})$ and $(\ref{6.18})$.
  Hence, the Abel map linearizes the divisor
  $\mathcal{D}_{ \hat{\mu}_0(x,t_r) \underline{\hat{\mu}}(x,t_{r})}$ on $\Omega$ with
  respect to $\tilde{x}, \tilde{t}_r$.
  \end{rem6.5}

\newtheorem{rem6.6}[lem6.1]{Remark}
  \begin{rem6.6}
    Remark $\ref{remark4.8}$ applies in the  present time-dependent
    context. Moreover, to obtain the theta function
    representation of $\psi_j$, $j=1,2,$, one can write $\widetilde{F}_{r}$
    in terms of $\Psi_k(\bar{\mu})$ and use $(\ref{5.46})$,
    in analogy to the stationary case discussed in Remark $\ref{remark4.9}$. Here
    we omit further details.
  \end{rem6.6}

At the end of this section, we turn to
the time-dependent
algebro-geometric initial value problem of HS2 hierarchy. We will show that
the solvability of the Dubrovin equations (\ref{5.56}) and
(\ref{5.57}) on $\Omega_\mu \subseteq \mathbb{R}^2 $ in fact implies equations
(\ref{5.3}) and (\ref{5.4}) on $\Omega_\mu$.

\newtheorem{the6.7}[lem6.1]{Theorem}
   \begin{the6.7}\label{theorem6.7}
    Fix $ n \in \mathbb{N}_0$,
     assume $(\ref{2.26a})$,
   and suppose that $\{\hat{\mu}_j\}_{j=0,\ldots,n}$
   satisfies the Dubrovin equations $(\ref{5.56})$, $(\ref{5.57})$
   on an open and connected set $\Omega_\mu \subseteq \mathbb{R}^2$,
   with  $\widetilde{F}_{r} (\mu_j)$ in  $(\ref{5.57})$
   expressed in terms of $\mu_k$, $k=0,\ldots,n$, by
   $(\ref{6.6})$.
      Moreover, assume that
   $\mu_j$, $j=0,\ldots,n,$ remain distinct and nonzero on $\Omega_\mu$.
     Then, $u, \rho \in C^\infty(\Omega_\mu)$, defined by
       \begin{equation}\label{6.26}
   u=-\frac{1}{2}\sum_{m=0}^{2n+1} E_m +\frac{1}{2}\sum_{j=0}^n \mu_j,
       \end{equation}
   and
      \begin{equation}
   \rho^2=u_x^2+2uu_{xx}- \frac{d^2}{dx^2}
    \Bigg(\Psi_2(\bar{\mu})-u\sum_{m=0}^{2n+1}E_m
    \Bigg), \label{6.26a18}
      \end{equation}
    satisfy the $r$th $HS2$ equation $(\ref{5.1})$, that is,
       \begin{equation}\label{6.27}
        \mathrm{HS2}_r(u,\rho)=0
        \quad \textrm{on \,$\Omega_\mu$},
       \end{equation}
     with initial values satisfying the $n$th stationary $HS2$ equation $(\ref{5.1a8})$.
   \end{the6.7}
\noindent
{\it Proof.}~Given the
  solutions $\hat{\mu}_j=(\mu_j,y(\hat{\mu}_j))\in
  C^\infty(\Omega_\mu,\mathcal {K}_n),$ $j=0,\ldots,n$
  of (\ref{5.56}) and (\ref{5.57}), we
  define polynomials $F_{n},$ $ G_n$, and $H_n$
  on $\Omega_\mu$ as in the stationary case (cf. Theorem \ref{theorem4.10})
  with properties
      \begin{align}
       &
        F_{n}(z)= \frac{1}{2}\prod_{j=0}^n (z-\mu_j), \label{6.28a1}\\
       &
        G_n(z)=\frac{1}{2}F_{n,x}(z), \label{6.28a2}\\
       &
       z^2G_{n,x}(z)=-H_n(z)-(\rho^2+u_{xx}z)F_{n}(z),\label{6.28a3}
          \\
       &
       H_{n,x}(z)=2(\rho^2+u_{xx}z)G_n(z), \label{6.28a4}\\
       &
       R_{2n+2}(z)=z^2G_n^2(z)+F_{n}(z)H_n(z),\label{6.28a5}
     \end{align}
treating $t_r$ as a parameter.
Define the polynomials $\widetilde{G}_r$ and $\widetilde{H}_r$  by
     \begin{align}
      &
        \widetilde{G}_r(z)=\frac{1}{2}\widetilde{F}_{r,x}(z)
        \quad \textrm{on $ \mathbb{C} \times \Omega_\mu$},\label{6.33} \\
     &
       \widetilde{H}_r(z)=-z^2\widetilde{G}_{r,x}(z)
       -(\rho^2+u_{xx}z)\widetilde{F}_{r}(z)
     \quad \textrm{on $ \mathbb{C} \times \Omega_\mu$},\label{6.34}
     \end{align}
respectively. Next, we claim that
\begin{equation}\label{6.35}
        F_{n,t_r}(z)=2(G_n(z)\widetilde{F}_{r}(z)-F_{n}(z)\widetilde{G}_r(z))
        \quad \textrm{on $ \mathbb{C} \times \Omega_\mu$}.
    \end{equation}
To prove (\ref{6.35}), one computes from (\ref{5.56}) and (\ref{5.57}) that
   \begin{align}
   &
   F_{n,x}(z)=-F_{n}(z) \sum_{j=0}^n \mu_{j,x} (z-\mu_j)^{-1},\label{6.36}
   \\
   &
    F_{n,t_r}(z)=-F_{n}(z) \sum_{j=0}^n
   \widetilde{F}_{r}(\mu_j) \mu_{j,x} (z-\mu_j)^{-1}.\label{6.37}
   \end{align}
Using (\ref{6.28a2}) and (\ref{6.33}), one concludes that (\ref{6.35}) is
equivalent to
    \begin{equation}\label{6.38}
    \widetilde{F}_{r,x}(z)=\sum_{j=0}^n (\widetilde{F}_{r}(\mu_j)
    -\widetilde{F}_{r}(z)) \mu_{j,x} (z-\mu_j)^{-1}.
    \end{equation}
Equation (\ref{6.38}) is proved in Lemma F.9 \cite{15}.
This in turn proves (\ref{6.35}).

Next, differentiating (\ref{6.28a2}) with respect to $t_r$  yields
    \begin{equation}\label{6.39}
    F_{n,xt_r}=2G_{n,t_r}.
    \end{equation}
On the other hand, taking the derivative of (\ref{6.35}) with respect to
$x$, and using (\ref{6.28a2}), (\ref{6.28a3}), (\ref{6.33}), one obtains
    \begin{align}\label{6.40}
     F_{n,t_rx}=&-2z^{-2}H_n\widetilde{F}_{r}
     -2(z^{-2}\rho^2+u_{xx}z^{-1})F_{n}\widetilde{F}_{r}+2G_n\widetilde{F}_{r,x}
       \nonumber \\
     &
     -2\widetilde{G}_{r,x}F_{n}-4\widetilde{G}_rG_n.
    \end{align}
Combining (\ref{6.28a2}), (\ref{6.33}), (\ref{6.39}), and (\ref{6.40}), one concludes
    \begin{equation}\label{6.41}
     z^2G_{n,t_r}(z)=\widetilde{H}_r(z)F_{n}(z)-H_n(z)\widetilde{F}_{r}(z)
      \quad \textrm{on $ \mathbb{C} \times \Omega_\mu$. }
    \end{equation}
Next, differentiating (\ref{6.28a5}) with respect to $t_r$, and using expressions
(\ref{6.35}) and (\ref{6.41}) for $F_{n,t_r}$ and
$G_{n,t_r}$, respectively, one obtains
    \begin{equation}\label{6.42}
        H_{n,t_r}(z)=2(H_n(z)\widetilde{G}_r(z)-G_n(z)\widetilde{H}_r(z))
        \quad \textrm{on $ \mathbb{C} \times \Omega_\mu$. }
    \end{equation}
Finally, taking the derivative of (\ref{6.41}) with respect to $x$,
and using expressions (\ref{6.28a2}), (\ref{6.28a4}), and (\ref{6.33}) for $F_{n,x}$,
$H_{n,x}$, and $\widetilde{F}_{r,x}$, respectively, one infers that
     \begin{equation}\label{6.43}
     z^2G_{n,t_rx}= F_{n}\widetilde{H}_{r,x}+2G_n\widetilde{H}_r
      -2(\rho^2+u_{xx}z)G_n\widetilde{F}_{r}
         -2H_n\widetilde{G}_r.
    \end{equation}
On the other hand, differentiating (\ref{6.28a3}) with respect to $t_r$,
using (\ref{6.35}) and (\ref{6.42}) for $F_{n,t_r}$ and
$H_{n,t_r}$, respectively, leads to
   \begin{align}\label{6.44}
    z^2G_{n,xt_r}&=2G_n\widetilde{H}_r-2H_n\widetilde{G}_r
    -(2\rho \rho_{t_r} + u_{xxt_r}z)F_{n}
  -2(\rho^2+u_{xx}z)(G_n\widetilde{F}_{r}-\widetilde{G}_{r}F_{n}).
   \end{align}
 Combining (\ref{6.43}) and (\ref{6.44}) then yields
   \begin{equation}\label{6.45}
   -2\rho \rho_{t_r} -u_{xxt_r}z-\widetilde{H}_{r,x}
  +2(\rho^2+u_{xx}z)\widetilde{G}_r=0.
   \end{equation}
Thus, we have proved (\ref{5.12})-(\ref{5.17}) and (\ref{5.38})-(\ref{5.40}) on $ \mathbb{C}
\times \Omega_\mu$ and hence conclude that
(\ref{6.27}) holds on $ \mathbb{C} \times \Omega_\mu$.
\quad $\square$

\newtheorem{rem6.8}[lem6.1]{Remark}
     \begin{rem6.8}
       Again we formulated Theorem $\ref{theorem6.7}$  in terms of
       $\{\mu_j\}_{j=0,\ldots,n}$ only. Obviously,
        the analogous result (and strategy proof) works
       in terms of $\{\nu_j\}_{j=0,\ldots,n}$.
     \end{rem6.8}

The analog of Remark \ref{remark4.12} directly extends to the current
time-dependent setting.

\section*{Acknowledgments}
The work described in this paper
was supported by grants from the National Science
Foundation of China (Project No.10971031; 11271079; 11075055),
Doctoral Programs Foundation of the Ministry of Education of China,
and the Shanghai Shuguang
Tracking Project (Project 08GG01).

\end{document}